\documentclass[%
 reprint,
 amsmath,amssymb,
 aps,
 prb,
floatfix,
]{revtex4-2}
\usepackage{graphicx}% Include figure files
\usepackage{dcolumn}% Align table columns on decimal point
\usepackage{bm}% bold math
\usepackage{placeins}
\usepackage[
  colorlinks=true,
  allcolors=blue
]{hyperref}
\usepackage[english]{babel}
\begin{document}
\preprint{APS/123-QED}

\title{Pressure induced electronic band evolution and observation of superconductivity in the Dirac semimetal ZrTe$_5$}
% Force line breaks with \\

\author{Sanskar Mishra$^{1}$, Nagendra Singh$^1$, Vinod K. Gangwar$^2$, Rajan Walia$^3$, Jianping Sun$^{4}$, Genfu Chen$^{4}$, Dilip Bhoi$^{5}$, Sandip Chatterjee$^{6}$, Yoshiya Uwatoko$^{5}$, Jinguang Cheng$^{4\dagger}$, Prashant Shahi$^{1}$}

 \altaffiliation{prashant.phy@ddugu.ac.in}%Lines break automatically or can be forced with \\
 \email{$^\dagger$ jgcheng@iphy.ac.cn}
 
\affiliation{
$^1$Department of Physics, Deen Dayal Upadhyaya Gorakhpur University, Gorakhpur, 273009, India\\$^2$Department of Physics, K.G.K. (P.G.) College, Moradabad, 244001, India\\$^3$Department of Physics, University of Allahabad, Prayagraj, 211002, India\\$^4$Beijing National Laboratory for Condensed Matter Physics and Institute of Physics,Chinese Academy of Sciences, Beijing, 100190 China\\$^5$Institute for Solid State Physics, University of Tokyo, Kashiwa, Chiba 277-8581, Japan\\$^6$Department of Physics, Indian institute of technology (IIT)-BHU, Varanasi, 221005, India}

\date{\today}% It is always \today, today,
\begin{abstract}
We report a comprehensive investigation of the pressure effects on the magnetotransport properties of the topological material ZrTe$_5$ within 1--8~GPa pressure range. With increasing pressure, the characteristic peak ($T_p$) in its electrical resistivity $\rho(T)$ first shifts to higher temperature and then moves quickly towards the lower temperature before disappearing eventually at 6~GPa. Beyond 6~GPa, the system exhibits metallic behavior across the entire temperature range, and superconductivity emerges below $T_c = 1.8$~K at 8~GPa. Based on the systematic magnetotransport measurement under pressure, we demonstrate that the superconductivity occurs following a significant electronic structure modulation possibly due to pressure induced structural changes near 6~GPa, which coincides with dramatic enhancement of the magnetoresistance (MR) reaching up to $\sim 1400\%$. Our experimental results are substantiated by density functional theory calculations as the application of pressure drastically alters the density of states near the Fermi level. Notably, multiple hole pockets emerge at the Fermi level from 4~GPa onward, and their contributions are further enhanced with increasing pressure. The combined experimental and theoretical investigation reveals a comprehensive evolution of electronic structure of Dirac semimetal ZrTe$_5$ under pressure and suggest a possible link between the Fermi surface reconstruction in the pressure range of structural transition and emergence of superconductivity.
\end{abstract}
\maketitle
%\tableofcontents
\section{\label{sec:level}	INTRODUCTION }
Pressure-driven superconductivity in topologically non-trivial materials continues to attract considerable attention, particularly for its feasibility in realizing topological superconductivity (TSC)\cite{Li2019}.Variety of topological materials have been explored in this context, including topological insulators (e.g., Bi$_2$Te$_3$, Bi$_2$Se$_3$) \cite{Czhang2011,Kirshenbaum2013,Zhu2013,JLZhang2011} and Dirac/Weyl semimetals (e.g., Cd$_3$As$_2$, WTe$_2$, MoTe$_2$) \cite{He2016,Kang2015,Qi2016}, which highlights the relevance of high-pressure studies in the search for unconventional superconducting states. In general, pressure-induced superconductivity in the topological systems is often a consequence of lattice or electronic instability induced by pressure, manifested as structural transitions or significant anomalies in the physical properties. For instance, superconductivity emerges in topological insulator Bi$_2$Te$_3$ upon compression up to 3.2 GPa, where the slope of Hall coefficient (dR$_H$/dP) exhibits a pronounced change \cite{Czhang2011,Kirshenbaum2013,Zhu2013,JLZhang2011}. In the Dirac semimetal Cd$_3$As$_2$, the structural transition takes place near 3 GPa, prior to the appearance of superconductivity at 8.5 GPa \cite{He2016}, and the symmetry breaking associated with the structural distortion has been proposed to be favorable for realizing topological superconductivity \cite{Kob2015}. The emergence of superconductivity in the Weyl semimetal WTe$_2$ at 10.5 GPa coincides with the complete suppression of its extreme magnetoresistance (XMR) and sign reversal of Hall coefficient \textit{$(R_H)$} \cite{Kang2015}. In contrast, superconductivity emerges in NbAs$_2$ before the XMR is completely suppressed, as the pressure modifies Fermi surface while maintaining the electron-hole compensation \cite{Li2018}. Furthermore, unconventional superconductivity has also been reported in several other topological semimetals such as LaBi \cite{Nayak2017} and MoP \cite{Lv2017,Chi2018}, for which electronic instabilities were found to be essential for realizing this state. Therefore, the application of high pressure can be employed as an effective approach to induce superconductivity by regulating the structural and/or electronic properties of topological materials. 

Zirconium pentatelluride ($\mathrm{ZrTe_5}$) is an established topological material with Dirac-like band dispersion \cite{Weng2014} and exhibits numerous unusual properties associated with its peculiar electronic states, including resistivity anomaly \cite{Okada1980,Jones1982}, chiral magnetic effect \cite{Li2016}, anomalous Hall effect (AHE) \cite{Liang2018}, anomalous thermoelectric effect \cite{zhang2019,Zhang2020}, magneto chiral anisotropy \cite{Wang2022} etc.~Previous studies have shown a significant effect of pressure on its electronic structure, leading to pressure-induced topological \cite{Fan2017,Liu2017,KK2024} and superconducting \cite{Zhou2016} transitions.~Recently, we investigated the thermoelectric and magneto-transport properties of ZrTe$_5$ single crystals under hydrostatic pressure up to 2 GPa \cite{Mishra2025}, and observed an upward shift of its characteristic resistivity peak towards higher temperatures with pressure.~The magnetotransport studies under hydrostatic pressure also reveals a phase change in quantum oscillations around  $\sim$ 2 GPa, suggesting pressure induced modification in Fermi surface topology similar to the report from other group \cite{dis2017}. Additionally, our previous study also revealed that the application of moderate pressure enhances simultaneously resistivity \textit{$\rho(T)$} and thermopower \textit{S(T)}, leading to a significant enhancement of the thermoelectric power factors at room temperature \cite{Mishra2025}.~These results clearly suggest that the application of pressure can significantly tune the electronic structure of ZrTe$_5$ even in low pressure regime (0-2 GPa).~High pressure studies reaching up to $\sim$ 50 GPa in $\mathrm{ZrTe_5}/ \mathrm{HfTe_5}$ have been reported by several groups, where a semimetal to superconducting transition associated with structural phase transition have been observed near 6 GPa \cite{Liu2017,Zhou2016,Qi_2016,Ana2020}, with the superconducting critical temperature \textit{T$_c$} increases up to $\sim$ 6 K at higher pressure.~These studies establish a link between the structural transition and superconductivity, though the precise pressure at which this transition occurs can vary between different crystal batches owing to the well-known sensitivity of the transport properties of ZrTe$_5$ to the crystal growth method.~While the past studies have primarily aimed at investigation of topological phase transition in low pressure regime or the observation of superconductivity at elevated pressures, a detailed understanding on evolution of electronic properties in the intermediate pressure range remain limited.~In particular, evolutions of MR, Hall resistivity ($\rho_{xy}$), resistivity peak (\textit{T$_p$}), and the electronic structure as function of pressure have not been adequately studied.

Therefore, in the present study, we have systematically investigated the magnetotransport properties of chemical vapor transport (CVT) grown ZrTe$_5$ single crystals under various hydrostatic pressures from 1 to 8 GPa in order to complement and extend the existing high pressure studies of transition metal pentatellurides (ZrTe$_5$). Our results confirmed a significant effect of pressure on the electrical transport properties of ZrTe$_5$. Notably, the characteristic resistivity peak (\textit{T$_p$}) displays a pronounced non-monotonic pressure dependence, which initially increases with pressure and then shifts towards lower temperature before ultimately vanishes near 6 GPa. The superconducting transition with \textit{T$_c$} =1.8 K was observed to occur at 8 GPa. The magnetotransport results (MR and Hall resistivity) provide systematic band evolution under pressure. The transition in bands from electron-like to hole-like began at 4 GPa, as evidenced by the emergence of significant non-linearity in MR and Hall data consistent with the DFT calculations. At 6 GPa, a perfect balance between the two type of carries is achieved marked by a dramatic increase in the low-temperature MR ($\sim$1400~\% at 8 T). In contrast with other topological materials, we observed distinctive non-monotonic pressure dependence of MR in ZrTe$_5$, e.g. MR first decreases sharply in the low-pressures regime (down to $\sim$50 \% at 2 GPa), recovers in the intermediate pressure range (200 \% at 3–5 GPa), and then rises sharply at 6 GPa ($\sim$1400 \%) along with the complete sign reversal of Hall resistivity. Our experimental observations are supported by first-principles DFT calculations, which display pressure-induced changes in calculated density of states (DOS) and band structures in the 1-6 GPa pressure range. Our results demonstrate that pressure plays a significant role in tuning properties of ZrTe$_5$, likely through a pressure-induced structural transition and/or associated electronic reconstruction.
                      
\begin{figure}
\includegraphics[width=\linewidth]{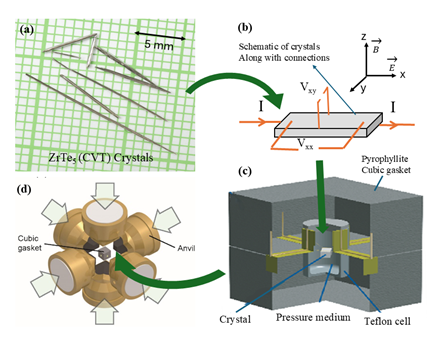}% Here is how to import EPS art
\caption{\label{fig:wide} (a) Optical image of CVT grown ZrTe$_5$ single crystals. (b) Illustration of four probe measurement of transport properties under applied magnetic field B. (c) Schematic views of the sample assembly inside the pyrophyllite cubic gasket. (d) Six anvils compressing the cubic gasket shown in figure (c). Panels (c,d) have been adapted from previous work of our group ref.\cite{Cheng2018,Uwatoko2008}.}
\label{fig1}
\end{figure}
\section{\label{sec:level}Experimental Details}

The ZrTe$_5$ single crystals used in this study were grown by the CVT technique \cite{Okada1980,Jones1982}. Structural and compositional characterizations together with the physical properties at the ambient pressure of these single crystals have been reported in our previous work \cite{Shahi2018}. A palm-type cubic anvil Pressure cell (PCAC) was used to measure magnetoresistance under hydrostatic pressures up to 8 GPa. The conventional four probe method was employed in resistivity measurement, where the direction of current was kept along ‘\textit{a}’ axis. In the magnetotransport measurements the direction of magnetic field was kept along ‘\textit{b}’ axis. For the generation of hydrostatic pressure, glycerol was used as the pressure-transmitting medium inside the pressure cell. Glycerol offers better hydrostaticity at low temperature in the intermediate pressure regime compared to other alternatives such as Daphene 7373 \cite{Osakabe_2008}. The pressure inside the cell was determined by the superconducting transition of lead (Pb) at low temperatures. Figure~\ref{fig1}(a) presents the optical image of the CVT grown ZrTe$_5$ single crystals which has been reproduced from our previous study \cite{Mishra2025}. Figure~\ref{fig1}(b) illustrates the schematic of four probe measurement configuration for the magneto-transport experiments under applied magnetic field B. Figure~\ref{fig1}(c, d) illustrates the pyrophyllite gasket along with the Teflon cell assembly in the center of PCAC, which are reproduced from our previous work \cite{Cheng2018,Uwatoko2008}.

The electronic band structure and density of states (DOS) were computed using Quantum espresso package \cite{Gian2017}. Generalized Gradient Approximation (GGA) was implemented using Perdew-Burke-Ernzerhof (PBE) functional \cite{Per2017} to account for exchange correlation effects. Additionally, we have also used Grimme’s D3 dispersion correction \cite{Grimme2010} to account for van der Waal’s interaction between different layers of ZrTe$_5$. The convergence criteria were set for the force threshold to 10$^{-5}$ atomic unit and for energy it is 10$^{-12}$ Ry. The pressure convergence was maintained within 0.05 kbar. For these calculations, scalar-relativistic ultra-soft pseudopotentials were used. Spin-orbit coupling (SOC) has been considered while calculating bands and DOS. Furthermore, band structure has been calculated taking a 15×15×4 k-point mesh, whereas a denser k-point mesh (double of the original one) has been used for non-self-consistent field (nscf) calculation to accurately map the Fermi level of the system.
\section{\label{sec:level}Results}
\subsection{\label{sec:level}Electrical resistivity under Pressure }
\begin{figure}
\includegraphics[width=\linewidth]{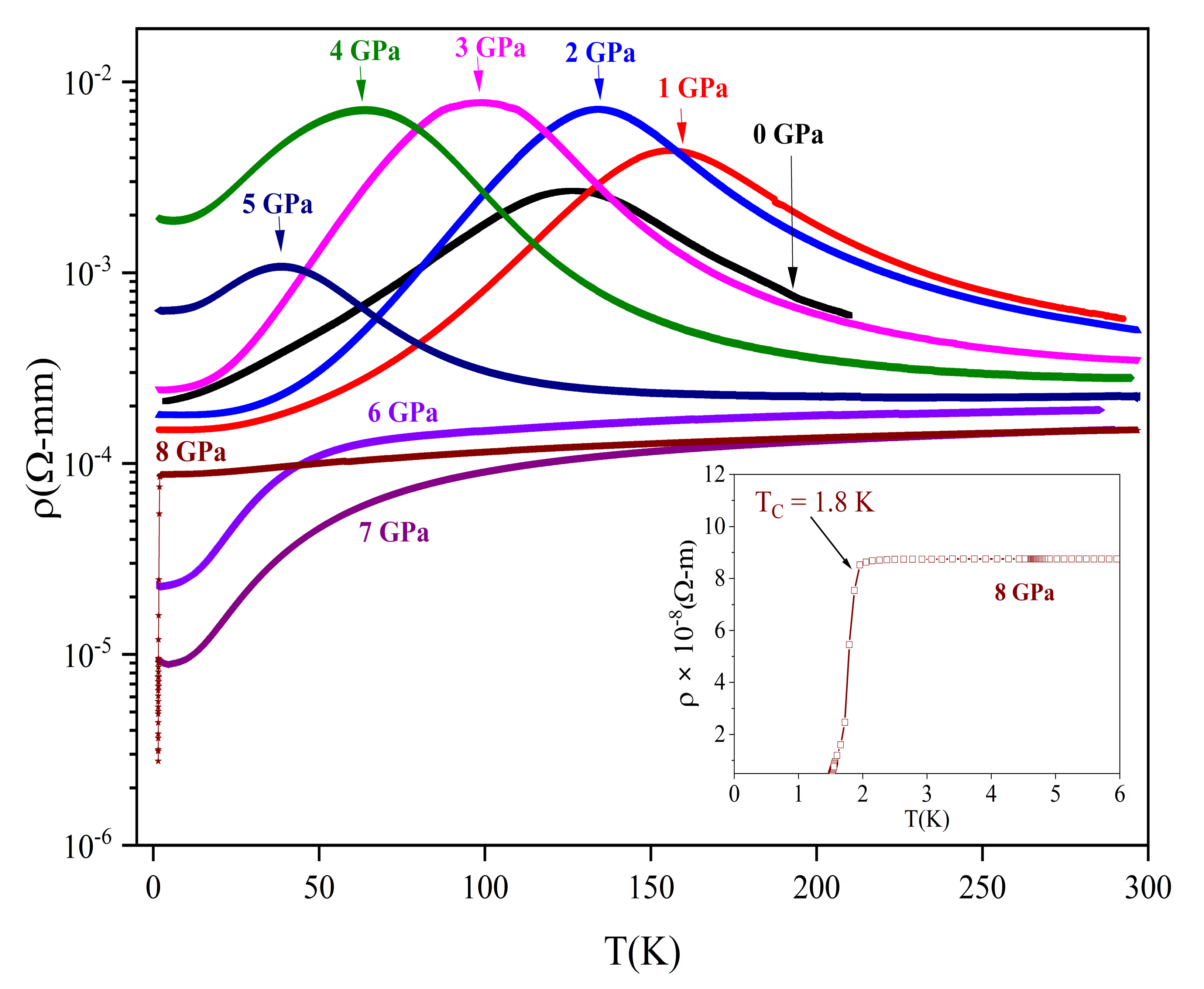}% Here is how to import EPS art
\caption{\label{fig:wide}~Temperature dependence of resistivity along a-axis of ZrTe$_5$ single crystals under various pressures up to 8 GPa. The application of pressure induces a notable change in resistivity peak (\textit{T$_p$} and \textit{$\rho$(T$_p$)}). The inset figure presents the low-temperature resistivity at 8 GPa, showcasing the superconducting transition at 1.8 K.}
\label{fig2}
\end{figure}

Figure~\ref{fig2} presents the temperature dependence of electrical resistivity \textit{$\rho(T)$} at various pressures $(0\leq P\leq8)$ for the ZrTe$_5$ crystals. At ambient pressure, $\rho(T)$ exhibits a broad peak at \textit{T$_p$}  $\sim$127 K. This resistivity peak is a characteristic of CVT-grown ZrTe$_5$ crystals and is found to be associated with sign change of the Hall coefficient and thermopower, indicates a change of the dominant charge carriers from electrons (n-type) to holes (p-type) \cite{Jones1982}. Using high-resolution angle-resolved photoemission spectroscopy (ARPES), previous studies have attributed this peak to a temperature-induced Lifshitz transition, where the electronic bands evolve from p-type semiconducting to n-type semimetallic behavior as the temperature decreases \cite{Zhang2017}.
With increasing pressure, \textit{T$_p$} initially shifts to higher temperature, reaching $\sim$155 K at 1 GPa. Beyond this pressure, \textit{T$_p$} decreases progressively, reaching as low as 38 K at 5 GPa, where a slight upturn like feature is observed below 15 K. Along with the downward shift of \textit{T$_p$}, \textit{$\rho$(T$_p$)} changes moderately up to 4 GPa and then decreases by about an order of magnitude at 5 GPa. The pressure-driven evolution of \textit{T$_p$} is consistent with previous studies on ZrTe$_5$ and its sister compound HfTe$_5$ \cite{Zhou2016,Piva2024}. The initial increase in \textit{$\rho$(T$_p$)} with pressure can be attributed to the enlarged band gap, while its subsequent decrease beyond 4 GPa signals band gap closure an interpretation supported by recent density functional theory calculation \cite{Gao2021}. At 6 GPa, the resistivity peak disappears completely, and a weak metallic character emerges in the temperature range of 50 $\leq$ T $\leq$ 300 K. Below $\sim$50 K, the resistivity exhibits a quick reduction, resulting in a residual resistivity $\rho$(1.5 K) approximately 10 times lower than that at ambient pressure. The drastic change in the $\rho$(T) curve profile at 6 GPa suggests significant alterations in the DOS near the Fermi level. At 7 GPa, the resistivity behavior is similar to that at 6 GPa, with only a slight reduction in overall magnitude. Upon further compression to 8 GPa, $\rho$(T) curve change to a typical metallic behavior with weak temperature dependence in a wide temperature range, followed by a sudden drop at T $\sim$1.8 K, inset of Fig.~\ref{fig2}, due to the occurrence of the superconducting transition, which aligns with previous reports in pentatelluride systems \cite{Liu2017,Zhou2016,Qi_2016}. 
 
\begin{figure*}
\includegraphics[width=1\linewidth]{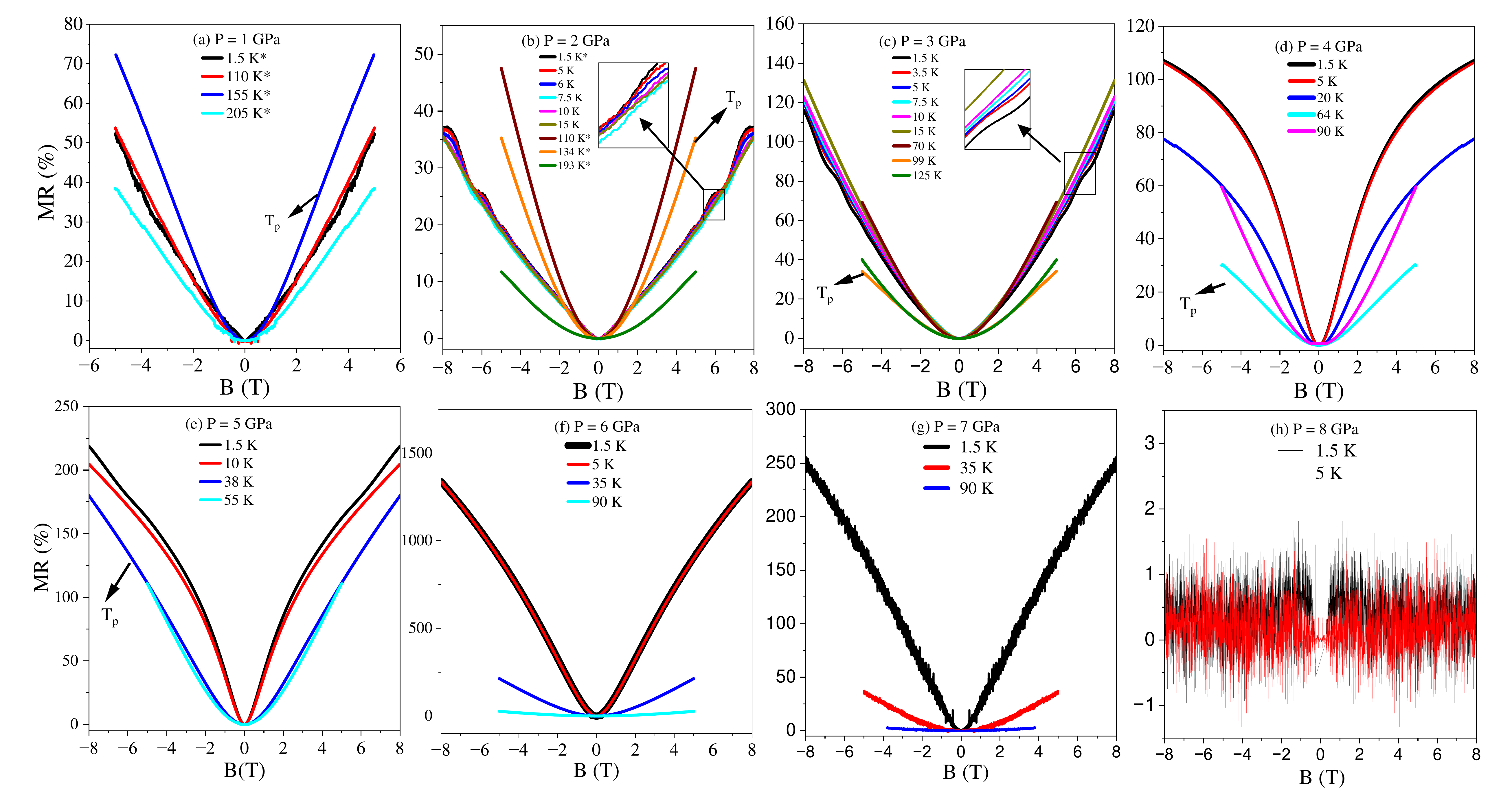}% Here is how to import EPS art
\caption{\label{fig:wide}~(a)–(h) Magnetic field dependence of magnetoresistance (MR \%) of ZrTe$_5$ at varying pressures (1–8 GPa). In panels (b) and (c), pronounced SdH oscillations are observed at low temperatures. The insets of panels (b) and (c) show magnified views of the overlapping data. At 8 GPa (panel h), the MR signal exhibits noticeable degradation, likely arising from experimental constraints associated with the palm cubic anvil cell at high pressure. Asterisks (*) next to temperatures at 1 and 2 GPa indicate data reproduced from our recent study \cite{Mishra2025}, included here for completeness.}
\label{fig3}
\end{figure*}   

\subsection{\label{sec:level}Magnetotransport properties under pressure}
\subsubsection{\label{sec:level}Magnetoresistance (MR) under pressure}
We investigated the evolution of longitudinal magnetoresistance (MR) of ZrTe$_5$ under varying pressure and temperature up to 8 GPa. The MR \% was derived from the longitudinal resistance (R$_{xx}$) using the conventional formula as-

\begin{equation}
  \mathrm{MR}\% = \left( \frac{R_{xx}(B) - R_{xx}(0)}{R_{xx}(0)} \right) \times 100\%
\label{eq1}
\end{equation}

where, $R_{xx}(B)$ and $R_{xx}(0)$ represent the longitudinal resistance in the presence and absence of a magnetic field, respectively. Figure~\ref{fig3} displays the MR \% as a function of magnetic field between 1--8~GPa pressure. At lower pressures ($P \leq 3~\text{GPa}$), the MR exhibits an almost linear, non-saturating field dependence, with a typical variation of $\mathrm{MR} \propto B^{1.3}$, as shown in Fig.~\ref{fig3} (a)--(c). However, upon increasing the pressure beyond 3~GPa, it begins to deviate from the linear nature, displaying a saturation-like tendency at higher magnetic fields, approximately following $\mathrm{MR} \propto B^{0.9}$ [Fig.~\ref{fig3}(d)--(f)].At 8 GPa [Fig.~\ref{fig3}(h)], the MR signal exhibits noticeable degradation, which may be attributed to the reduced signal magnitude under extreme pressure conditions and possible experimental constraints related to the high-pressure cell or the pressure-transmitting medium.

\begin{figure*}
\centering
\includegraphics[width=1\linewidth]{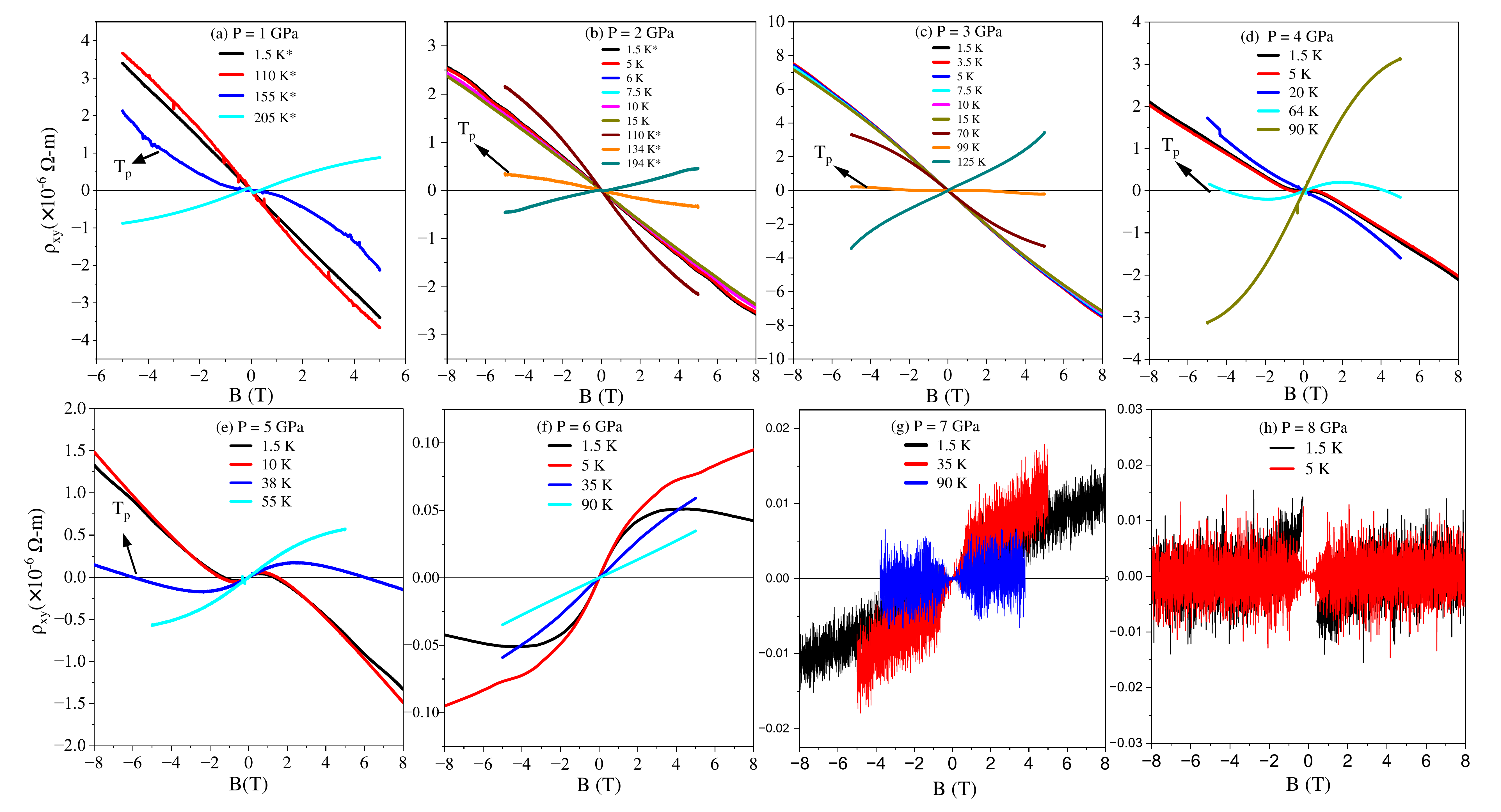}% Here is how to import EPS art
\caption{\label{fig:wide}(a)–(h) Magnetic field dependence of the Hall resistivity ($\rho_{xy}$) of ZrTe$_5$ under varying pressures up to 8 GPa. In the low-pressure regime (1–3 GPa), $\rho_{xy}$ varies nearly linearly with magnetic field (a–c). In the intermediate-pressure regime (4–6 GPa), $\rho_{xy}$ exhibits pronounced non-linear behavior, accompanied by a complete sign reversal at 6 GPa (d–f). In the high-pressure regime (7–8 GPa), the $\rho_{xy}$ signal shows noticeable degradation due to the reduced Hall voltage under extreme pressure conditions (g–h); nevertheless, the Hall sign reversal clearly persists up to 7 GPa. Asterisks (*) next to temperatures at 1 and 2 GPa indicate data reproduced from our recent study \cite{Mishra2025}.}
\label{fig4}
\end{figure*}

At lower field in the 4-6 GPa pressure range, the variation of MR is much steeper with cusp V-shaped features below \textit{T$_p$}, which may tentatively be associated with spin dependent scattering mechanism as pressure may strengthen the spin orbit interaction. Remarkably, at 6 GPa there is dramatic enhancement in MR reaching as high as $\sim$1400 $\%$ at low temperature and 8 T magnetic field [Fig.~\ref{fig3}(f)]. Such a huge MR implies a substantial modification in the electronic band structure of the system. As the pressure is raised, the MR near \textit{T$_p$} also varies systematically, i.e. it is the highest at 1 GPa, intermediate at 2 GPa and becomes significantly suppressed beyond 3 GPa. A comparison to our previous ambient pressure study of ZrTe$_5$ samples of the same batch \cite{Shahi2018} reveals a significant suppression of MR by pressure, similar as those reported by other groups \cite{dis2017}. It should be noted that at pressures of 2 and 3 GPa, pronounced Shubnikov de Haas (SdH) oscillations can be observed in MR curves at low temperatures, which we have discussed in detail in our recent study \cite{Mishra2025}.

\subsubsection{\label{sec:level} Hall Resistivity $(\rho_{xy})$ under pressure}
\begin{figure*}
\includegraphics[width=0.85\linewidth]{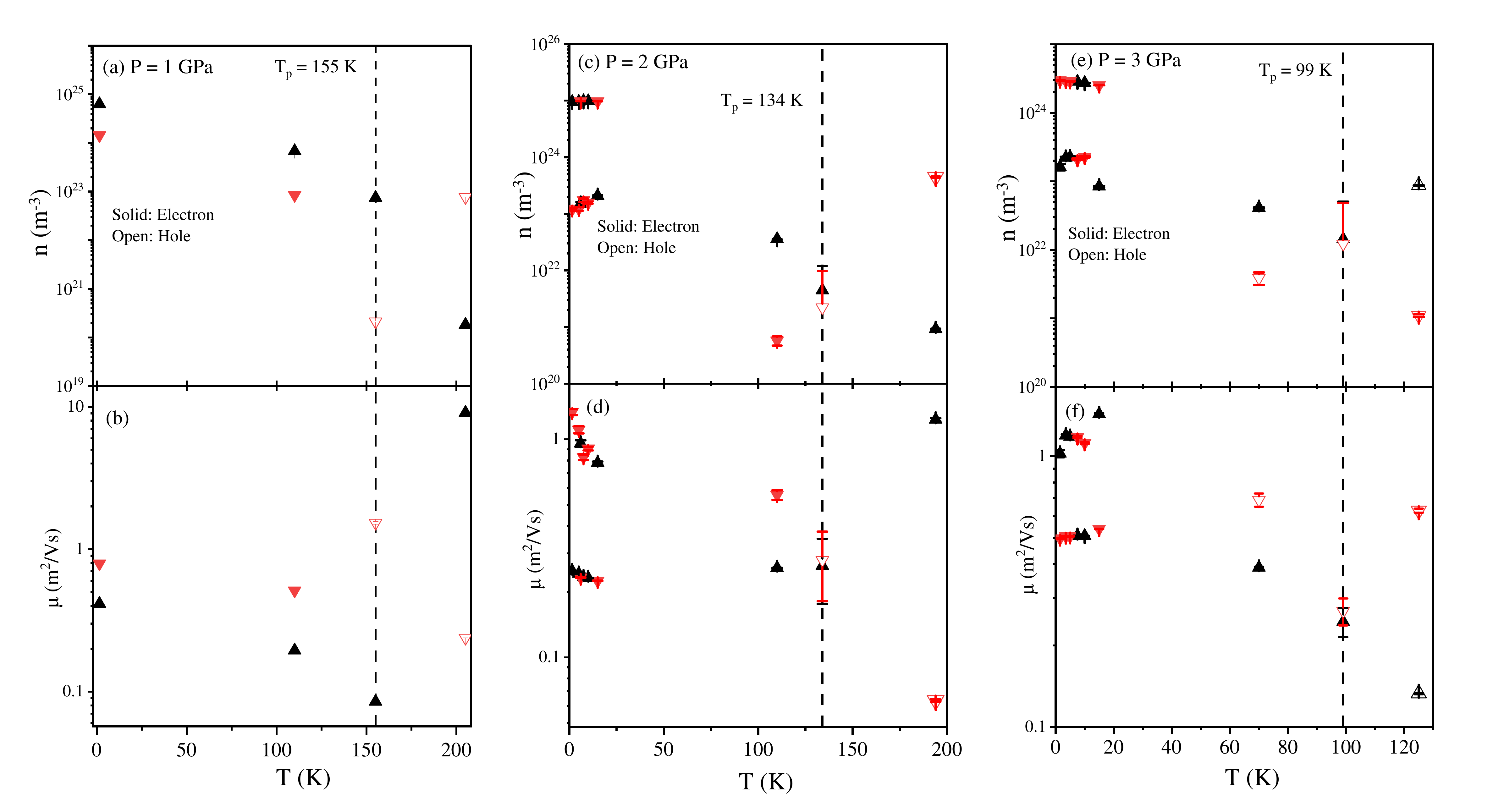}% Here is how to import EPS art
\caption{\label{fig:wide} Temperature dependence of carrier density (\textit{n}) (a, c, e) and mobility (\textit{µ}) (b, d, f) at 1-3 GPa pressure. Here, the solid symbols have been used to denote electron while the open symbols represent holes.}
\label{fig5}
\end{figure*}

\begin{figure*}
\includegraphics[width=0.875\linewidth]{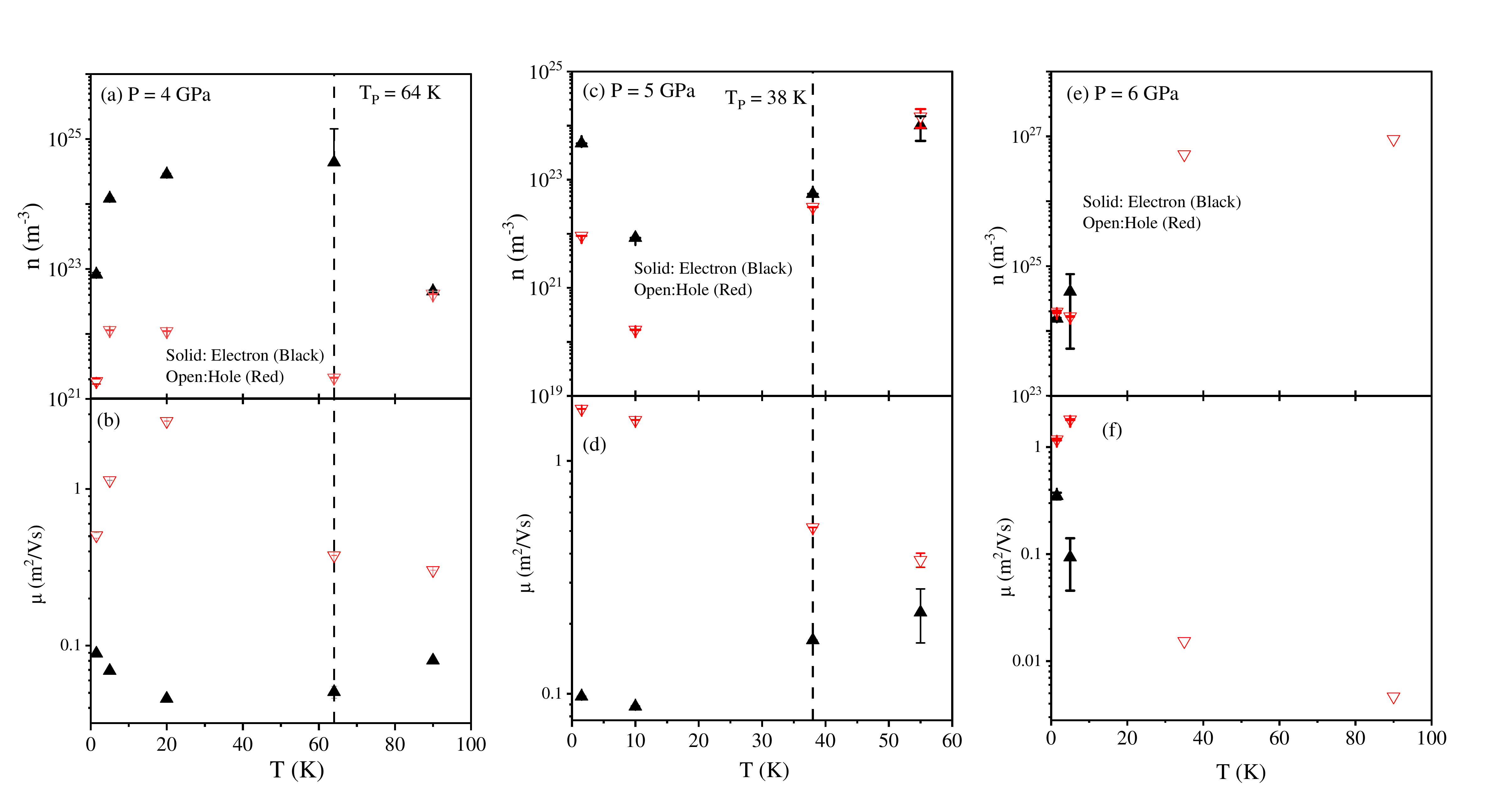}% Here is how to import EPS art
\caption{\label{fig:wide} Temperature dependence of carrier density (\textit{n}) (a, c, e) and mobility (\textit{µ}) (b, d, f) at 4-6 GPa. Here, the solid symbols have been used to denote electron while the open symbols represent holes.}
\label{fig6}
\end{figure*}

\vspace{-3mm}
To obtain the carrier information involved in pressure-dependent transport phenomenon, we measured the Hall resistivity ($\rho_{xy}$) as functions of magnetic field at various temperatures and pressures [Figure~\ref{fig4}]. Up to 5 GPa, where \textit{T$_p$} exists, the observed negative slope at \textit{$T<T_p$} of \textit{$\rho_{xy}$} reflects the dominance of electrons as the majority carrier. However, above T$_p$, the slope changes to positive, indicating a switch to hole dominated charge transport. At \textit{T$_p$}, where both electron and hole coexist, the \textit{$\rho_{xy}$} curve becomes relatively flat, highlighting the contribution of both carrier types. In 4-6 GPa pressure range, a pronounced curvature in \textit{$\rho_{xy}$(B)} emerges and this feature becomes prominent with pressure. Remarkably, the initial slope of  \textit{$\rho_{xy}$} (B) at low temperatures changes from negative to positive between 5 and 6 GPa, signaling a switch in the dominant carrier type from electron to hole-like.
Furthermore, based on the increasing non-linearity in MR and Hall, we estimated the charge carrier density (n) and mobility (µ) by fitting the Hall conductivity (\textit{$\sigma_{xy}$}) using two band model, given in Figure~\ref{fig5} $\&$ \ref{fig6}. Here, the Hall conductivity can be calculated from the resistivity tensor as $\sigma_{xy} = \frac{\rho_{xy}}{\rho_{xx}^2 + \rho_{xy}^2}$, where \textit{$\rho_{xx(xy)}$} is the longitudinal (Hall) resistivity. Using these values, the complete formula for Hall conductivity becomes- 
\begin{equation}
\sigma_{xy} = \pm eB \left( 
\frac{n_{1} \mu_{1}^{2}}{1 + \mu_{1}^{2} B^{2}} 
+ 
\frac{n_{2} \mu_{2}^{2}}{1 + \mu_{2}^{2} B^{2}} 
\right)
\label{eq2}
\end{equation}

Here, \textit{n$_1$} and \textit{n$_2$} are the density, \textit{$\mu_1$} and \textit{$\mu_2$} are the mobility of two different types of carriers, and ‘\textit{e}’ is fundamental electronic charge, and B is the applied magnetic field taken as independent variable. For each temperature and pressure, a magnetic field window 0-5 T was used to ensure the consistency in the procedure at for all measurement. The quality of these fits can be assessed via the $\sigma_{xy}$ vs $B$ curves provided in the Appendix~\ref{appC} for all pressures at 1.5 K. The negative or positive sign is considered depending on whether the corresponding carriers are electrons or holes, respectively.\\
Figure~\ref{fig5} and \ref{fig6} represents the estimated carrier density (n) and mobility (\textit{$\mu$}) at 1-3 and 4-6 GPa pressures respectively. The corresponding data at pressure 7 and 8 GPa is not shown here due to significant deterioration in singnal quality. It is clear from Figure~\ref{fig5} that in the low-pressure regime (1-3 GPa) the conduction is predominantly electronic with carrier density n $\sim$ 10$^{23}$-10$^{24}$ m$^{-3}$ below the peak temperature (\textit{$T_p$}), while both electrons and holes participate in conduction near \textit{$T_p$}, albeit with slightly lower densities. Above \textit{$T_p$}, the hole density increases significantly and becomes nearly two order higher than that of electrons. At each pressure there is substantial redistribution of carries is observed with increasing temperature, consistent with the temperature-induced Lifshitz transition as observed at ambient pressure \cite{Zhang2017,Y2017}. The higher mobility of electrons at 2 and 3 GPa leads to the pronounced Shubnikov-de Haas (SdH) quantum oscillations in MR as seen in Fig.~\ref{fig3} (b,c).\\

Upon further increasing pressure beyond 3 GPa the hole dominated conduction becomes more prominent. As shown in Figure~\ref{fig6}, the higher mobility holes start appearing right from low temperature, in contrast with results shown in Figure~\ref{fig5}. This observation suggests that the phenomenon of bipolar conduction which was typically confined near \textit{T$_p$} at low pressures low (1-3 GPa, Fig.~\ref{fig5}) persists throughout the entire temperature range because of pressure induced band modulations, and this extended bipolar conduction seems to be responsible for the enhanced  $\rho$(1.5K)  at 4 and 5 GPa [Figure~\ref{fig2}].
Furthermore, Figure~\ref{fig6} shows that the contribution from holes keeps increasing with temperature at each pressure. In between 5 and 6 GPa, the band becomes increasingly hole-like as evidenced by the positive slope of $\rho_{xy}$ at 6 GPa pressure [Fig.~\ref{fig4}(f)]. At 6 GPa, the electrons with comparable density as holes appear at low temperatures, which causes the dramatic enhancement in MR $\sim 1400 \%$ at magnetic field of 8 T. In addition, due to the linearity of $\rho_{xy}$ and MR at T = 35 K and 90 K two carriers model failed to fit the Hall conductivity reliably. The linearity of $\rho_{xy}$ at 35 and 90~K (6~GPa) indicates that transport at these temperatures is dominated by a single type of charge carrier, namely holes. Multiple bands may still exist, but their contributions to the Hall signal are negligible. Thus, we applied one band model viz. $n = \frac{dB}{|e| \, d\rho_{xy}}$ and $\mu = \frac{1}{|e| \, n \rho_{xx}}$ which indicates the hole dominated transport with highest carrier density throughout the investigated pressure. 
\subsection{\label{sec:level} Electronic band structure and Density of states (DOS) under pressure}

\begin{figure*}
\includegraphics[width=0.85\linewidth]{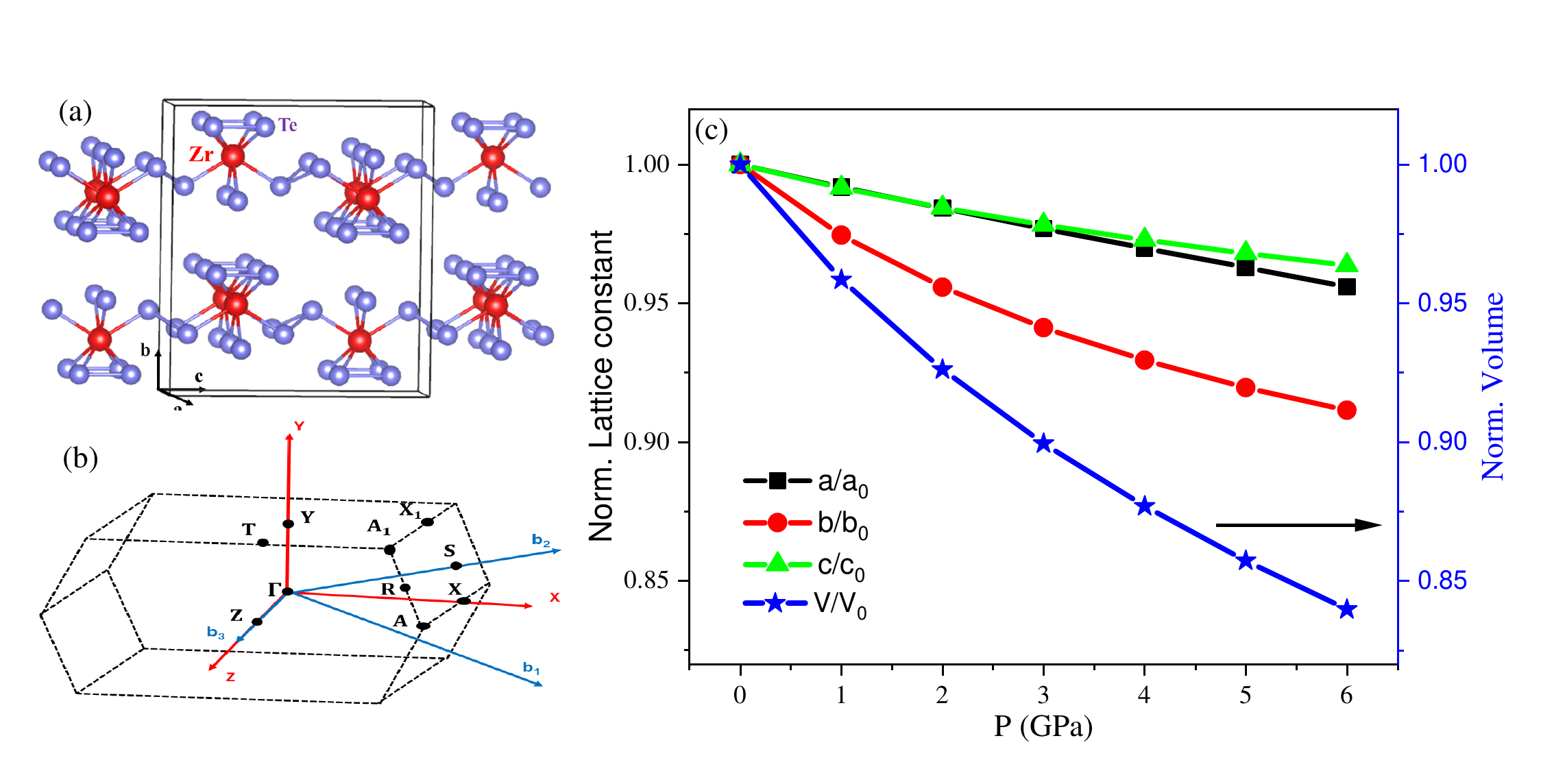}% Here is how to import EPS art
\caption{\label{fig:wide}(a) Crystal structure of ZrTe$_5$ in which trigonal chains of ZrTe$_3$ run along the a axis and different chains are joined along the c axis by zig-zag Te atoms. In ac plane it forms 2D ZrTe$_5$, which are stacked along \textit{b} axis by weak van der Waals forces.(b)  Brillouin zone of ZrTe$_5$ with the high symmetry points along which band structure has been calculated.(c) Pressure dependent normalized lattice constants and volume of ZrTe$_5$ unit cell.}
\label{fig7}
\end{figure*}

In Fig.~\ref{fig7}(a), we show the two layers of ZrTe$_5$ extended in the \textit{ac}-plane, which are stacked along the \textit{b}-axis by weak van der Waals interactions, forming a 3D ZrTe$_5$ crystal. The interlayer bonding energy is as low as that between the layers of graphite~\cite{Weng2014}; therefore, it is highly sensitive to external perturbations such as pressure and strain. The Brillouin zone along with the high-symmetry points is shown in Fig.~\ref{fig7}(b), where $\mathbf{b}_1$, $\mathbf{b}_2$, and $\mathbf{b}_3$ are the reciprocal lattice vectors. Figure~\ref{fig7}(c) displays the pressure dependence of the optimized lattice constants ($a$, $b$, and $c$) and the unit-cell volume from ambient to 6~GPa pressure. The calculated ambient-pressure lattice parameters, $\textit{a} = 4.03476\,\text{\AA}$, $\textit{b} = 14.87163\,\text{\AA}$, and $\textit{c} = 13.6695\,\text{\AA}$, are in good agreement with the experimentally reported values ($\textit{a} = 3.9830\,\text{\AA}$, $\textit{b} = 14.87163\,\text{\AA}$, and $\textit{c} = 13.6695\,\text{\AA}$)~\cite{Shahi2018}. As we increase the pressure, all lattice parameters decrease unevenly, with the order $\textit{b} > \textit{a} > \textit{c}$. Since the \textit{b}-axis represents the interlayer spacing, pressure reduces the interlayer spacing of ZrTe$_5$ crystals to about 7 \% of the original value at 6~GPa, which is also consistent with previously reported DFT calculations~\cite{Gao2021}. The calculated unit-cell volume shows a similar linear dependence on pressure as the lattice constants. At 6~GPa, it remains at approximately 85 \% of its original value.

 \begin{figure*}
\includegraphics[width=0.85\linewidth]{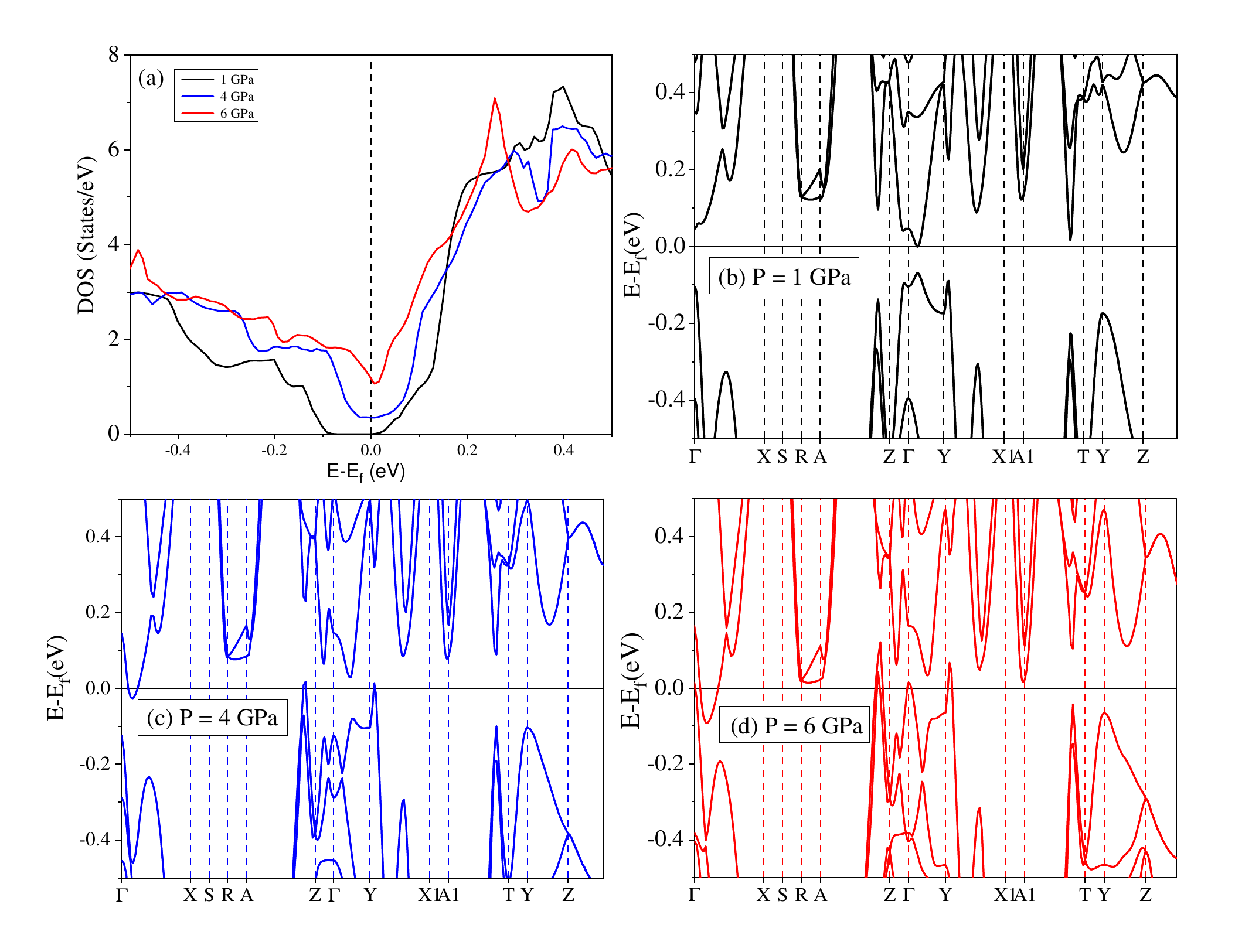}% Here is how to import EPS art
\caption{\label{fig:wide}(a)Electronic density of states (DOS) at selected pressure of 1, 4 and 6 GPa. (b)–(d) Band structure at pressures of 1, 4, and 6 GPa, respectively. Figure 8(b) has been reproduced from our previous study \cite{Mishra2025}}
\label{fig8}
\end{figure*}
We computed the electronic band structure and density of states (DOS) by optimizing the atomic positions and lattice parameters of ZrTe$_5$ at different pressures up to 6 GPa [Figure~\ref{fig8}].The spin-orbit coupling (SOC) effects were considered for all these calculations.Figure~\ref{fig8}(a) illustrates the density of states (DOS) at selected pressures of 1, 4 and 6 GPa.It is evident from this figure that as the pressure increases, there is a significant change in the DOS at the Fermi level notably, the states below the Fermi level are populated. Figures~\ref{fig8}(\textit{b–d}) show the band structures at 1, 4, and 6 GPa.The ambient pressure electronic bands and DOS have been discussed in detail in our recent study \cite{Mishra2025} and the data corresponding to 2, 3 and 5 GPa are given in Appendix~\ref{appD}.\\
At ambient pressure, the valance band maxima (VBM) and conduction band minima (CBM) meet at the Fermi level forming {Dirac} like band dispersion at the $\Gamma$-point \cite{Mishra2025}. At 1 GPa, there is an opening of the band gap at the ${\Gamma}$ point. In this case, the CBM and VBM shifts along {$\Gamma$-Y} direction with \textit{$\Delta\sim$}70 meV, Fig.~\ref{fig8}(b). As the pressure increases, the band gap persists up to 3 GPa. However, the indirect band gap decreases because additional Fermi pockets approach the Fermi level. The band structure at 4 GPa shown in Fig.~\ref{fig8}(c) illustrates that two-hole pockets cross the Fermi level near the \textit{$\Gamma$}-point (\textit{Z-A}, \textit{Y}-point) and one electron pocket crosses the Fermi level along the \textit{$\Gamma$-X} direction.The band structure at 6 GPa, Fig.~\ref{fig8}, indicates that the bands become more hole-like compared to the lower pressures as the extent of the Fermi pockets increases at the Fermi level.\\

\section{\label{sec:level} Discussion}
\begin{figure*}
\includegraphics[width=0.85\linewidth, height=8cm]{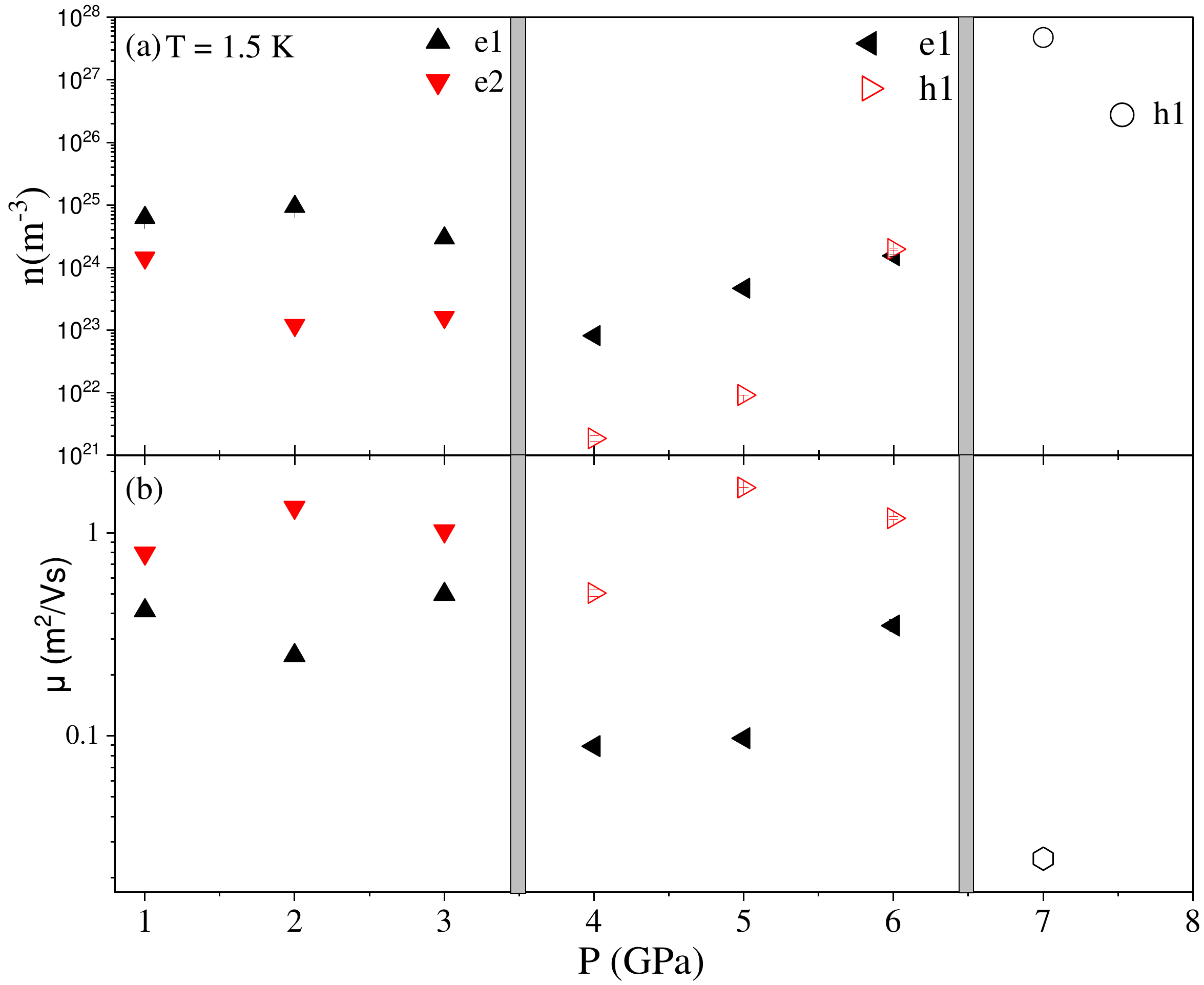}% Here is how to import EPS art
\caption{\label{fig:wide}(a)Density and (b) mobility of carriers as function of pressure at 1.5 K, extracted by fitting of Hall conductivity using the one-band or two-carrier models.}
\label{fig9}
\end{figure*}

The effect of pressure on the electronic properties of transition metal pentatellurides (ZrTe$_5$/HfTe$_5$) manifests a rich interplay between the band topology, carrier dynamics and superconductivity consistent with the general trend observed in other well-known topological materials like Bi$_2$Te$_3$, Cd$_3$As$_2$, WTe$_2$ etc\cite{Zhu2013,JLZhang2011,Kang2015,Qi2016,Kob2015,Chi2018}. Superconductivity in these materials generally coincides with or triggered by the instabilities in its electronic or lattice structure. In the present work, our comprehensive transport and computational results for the CVT grown ZrTe$_5$ single crystals displays a pressure induced transition, leading to pronounced Fermi surface reconstruction and ultimately the superconductivity. At ambient pressure, ZrTe$_5$ displays a characteristic peak in resistivity consistent with the previous reports\cite{Shahi2018}. However, on application of pressure we observe a significant non-monotonic change in position and magnitude of resistivity peak. More importantly, in between 5 to 6 GPa pressure, the resistivity peak disappears completely indicating the pressure induced reconstruction of underlying electronic band structure. The concomitant change in residual resistivity under pressure, further highlights the impact of high pressure on scattering phenomenon.The immediate evidence of the Fermi surface reconstruction comes from our pressure dependent magnetotransport experiments (MR and Hall resistivity, Fig.~\ref{fig13} $\&$~\ref{fig14}). There is a huge suppression in MR compared to our ambient pressure study of same samples\cite{Shahi2018}. Approximately linear field dependence of MR and Hall resistivity in low pressure regime (1-3 GPa) points to a simple, electron dominated transport, while in the intermediate pressure regime (4-6 GPa) non-linearity emerges in both MR and Hall resistivity indicates the participation of holes along with the electrons i.e. multi-band transport. This feature is quantitatively confirmed by the analysis of carrier density (\textit{n}) and mobility (\textit{µ}) extracted from two carrier model [Eq.~\ref{eq2}] revealing a systematic pressure induced increase of hole concentration. The electronic structure of ZrTe$_5$ is known to host multiple electron and hole pockets, especially under applied pressure~\cite{KK2024}, but here we employ a simplified two-band model as an effective empirical description of the magnetotransport. This approach is commonly used to identify the dominant conduction channels in topological semimetals~\cite{Eguchi}, where it reliably capture the essential features of Hall and magnetoresistance data even in the presence of multiple bands. The extracted carrier densities and mobilities thus represent effective transport parameters explaining the pressure-induced evolution of the system. It is clear from Figure~\ref{fig9} that this system evolves from the electron dominated band at low temperature to the nearly perfect electron hole compensated band at 6 GPa, causing the extreme magnetoresistance (XMR) $\sim$ 1400 \%. The sign reversal of the Hall resistivity, confirms a crossover from electron to hole dominated band beyond 6 GPa, with linear positive slope is recovered by 7 GPa [Fig.~\ref{fig5}(g)].\\
Our experimental observations are well supported by first-principles density functional theory (DFT) calculations. At ambient pressure, the conduction and valence bands nearly touch at the $\Gamma$ point, producing a {Dirac-like} dispersion as a result of an accidental Dirac semimetal state\cite{Mishra2025,Fan2017,Man2016}, in agreement with the observed non-trivial Berry phase reported previously by us and other groups \cite{Mishra2025,dis2017}. In low pressure regime (1–3 GPa), the calculations reveal a small band gap opening (Band wrapping) at $\Gamma$ point, consistent with the resistivity enhancement. More critically, on further raising pressure the calculation results in increased density of states at the Fermi level along with the emergence of multiple Fermi pockets origination at the $\Gamma$ point, along the $\Gamma$-X, A-Z and Y-X directions. The systematic band evolution in this study supports a critical change in the electronic band structure of ZrTe$_5$ under pressure, explaining the multiband conduction and providing the relation between the electronic structure modification and experimentally observed phenomenon. This dramatic electronic instability under pressure may be intimately linked to a structural transformation reported by Zhou et al. near 6 GPa \cite{Zhou2016}. Beyond this pressure, we expect the system to enter into new high-pressure phase which triggers the finally observed phenomenon i.e. the superconductivity at 8 GPa with \textit{T$_c$} $\sim$ 1.8 K. Here, we observed the emergence of superconductivity 8~GPa, which differs nearly by 2~GPa from the critical pressure for superconductivity as reported by Zhou \textit{et al.}($\sim$6~GPa)~\cite{Zhou2016}. This discrepancy likely arises from the combination of experimental and intrinsic sample dependent factors. First, in both the cubic anvil cell (CAC) and diamond anvil cell (DAC), the applied pressure typically decreases upon cooling. Therefore, a pressure difference of approximately $\sim$2~GPa is plausible, considering that Zhou \textit{et al.} performed pressure calibration at room temperature, whereas our calibration was carried out at low temperature. In addition, the use of different pressure-transmitting media (glycerol versus Daphne~7373), each with distinct hydrostatic limits, may introduce a minor calibration offset. Apart from this, ZrTe$_5$ is highly sensitive to synthesis conditions, as evidenced by the wide variation in ambient-pressure transport properties reported across different studies~\cite{Shahi2018,Mishra2025}. To further examine this aspect, we measured the temperature-dependent resistivity ($\rho$--$T$) of flux-grown ZrTe$_5$ crystals under conditions similar to those used in the present work. These measurements reveal a superconducting transition near 7~GPa with a critical temperature $T_c \approx 3.4$~K (Appendix~A), highlighting the crucial role of sample quality in stabilizing superconductivity. Taken together, these factors are likely relevant in accounting for the observed discrepancy in the superconducting transition temperature between the two studies.
 In Figure~\ref{fig3} \&~\ref{fig4}, the magnetotransport results at 7--8 GPa pressure is presented. The deterioration of the magneto transport signal quality after 6 GPa may likely be a consequence of structural or superconducting transition; however, the similarity between 1.5 K and 5 K data at 8 GPa rules out this possibility. We therefore believe it is due to experimental limitation of high pressure cell or the pressure transmitting medium. The emergence of superconductivity in the vicinity of the structural phase transition strongly suggests an intimate correlation between the pairing mechanism and underlying electronic structure.\\
\section{\label{sec:level} Conclusion}
In the present work, we have studied the effect of high pressure on the magnetotransport properties of CVT-grown ZrTe$_5$ single crystals. Resistivity measurements under varying pressures (0-8 GPa) reveal a non-monotonic and significant shift in the resistivity peak (T$_p$) and its eventual disappearance at 6 GPa. A superconducting transition is observed at 8 GPa with T$_c$=1.8 K. Our experimental results reveal a pressure-induced electronic transition, as evidenced by drastic change in transport and magnetotransport properties near 6 GPa. Between 4- 6 GPa, the hole density increases while the electron density decreases, which suggests the pressure induced reconstruction of electronic bands. At 6 GPa, a nearly balanced carrier population potentially explains the emergence of a dramatic MR of $\sim$1400 \% at 8~T. Additionally, complementary density functional theory calculations support our experimental findings, showing a substantial pressure-induced change in the density of states, leading to the emergence of multiple electron and hole pockets at the Fermi level.These findings suggest that pressure plays a significant role in driving the electronic transition along with the structural transition, and promoting the emergence of superconductivity in ZrTe$_5$. These observations deepen our understanding of pressure regulation on topological systems and provide a pathway for exploring novel quantum states.

\section{\label{sec:level}Acknowledgments}
S.M. acknowledges the support from the Science and Engineering Research Board (SERB), India, for the Junior Research Fellowship (File No.-SRG/2019/001187). V.K.G. acknowledges to the SERB India for the award of the SERB International Research Experience (SIRE) fellowship (File No.- SIR/2022/000804). P.S. acknowledges to the SERB India for the award of the Start-up research grant (SRG/2019/001187), University Grant Commission (UGC) for Basic Scientific Research (BSR) fund (UGC File No.- 30-505/2020(BSR)), UGC-DAE Centre for providing the fund (CRS/2021-22/01/415) and SERB India for SERB-SIRE fellowship (File No.- SIR/2022/000752). J.G.C. is supported by the National Key Research and Development Program of China (2023YFA1406100, 2021YFA1400200), the National Natural Science Foundation of China (12025408, U23A6003), and CAS PIFI program (2024PG0003).

\section{Appendixes}
\appendix
\section{}
\subsection*{ Resistivity Measurements of Flux-Grown ZrTe$_5$ at 1 and 7~GPa}
\label{appA}
\begin{figure}
\includegraphics[width=1\linewidth]{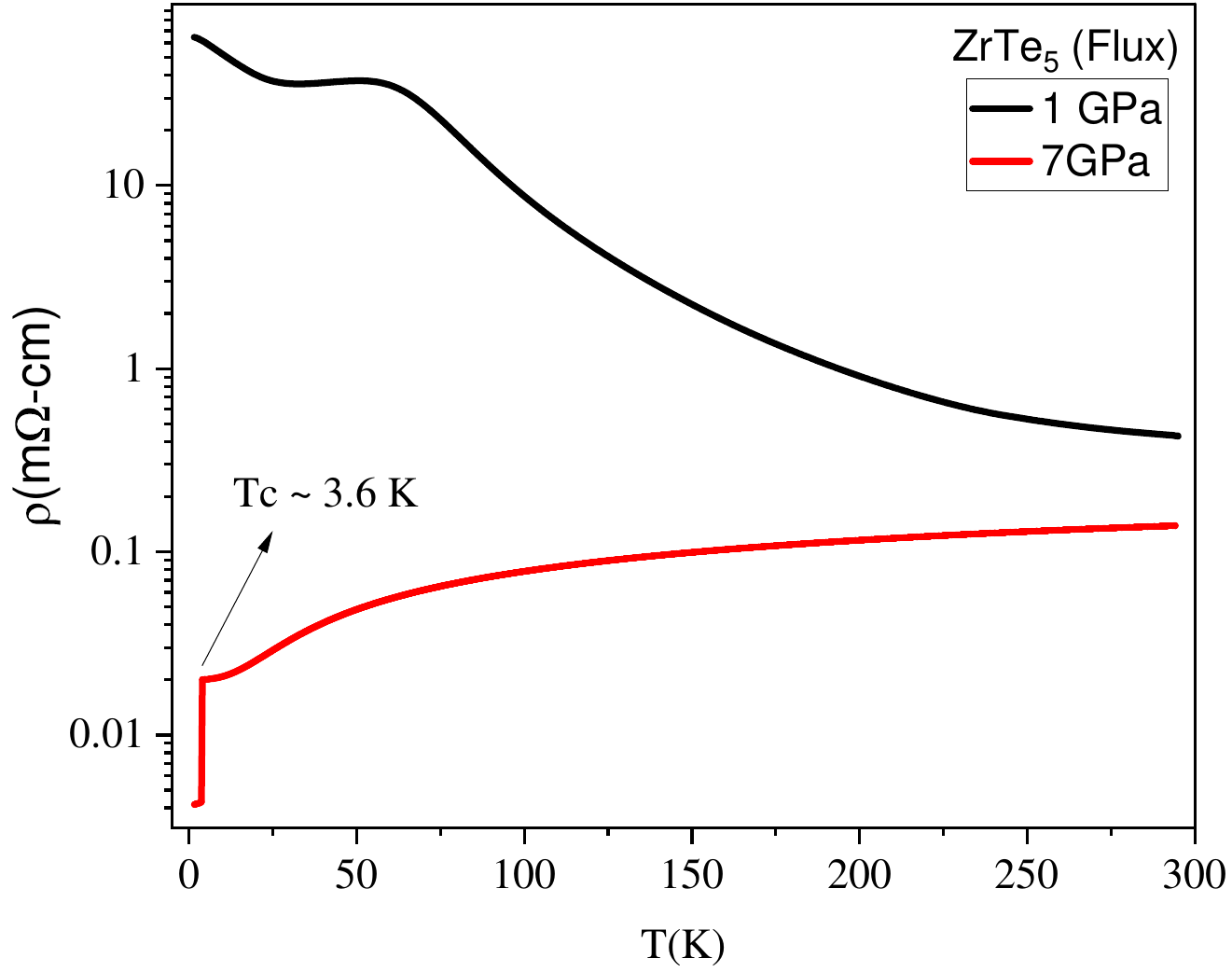}% Here is how to import EPS art
\caption{\label{fig:wide}(Appendix~\ref{appA}) Pressure-dependent resistivity as a function of temperature for flux-grown ZrTe$_5$ at 1 and 7~GPa}
\label{fig10}
\end{figure}

To support the claim that the superconducting critical pressure ($P_c$) in ZrTe$_5$ crystals varies between different crystal batches, we present complementary resistivity data obtained from flux-grown ZrTe$_5$ crystals [Fig.~\ref{fig10}]. The details of synthesis and structural parameters has been reported in Shahi \textit{et al.}~\cite{Shahi2018}. The measurements were carried out under identical experimental conditions to those described in the main text.
\section{}
\subsection*{ Raw magnetotransport data and symmetrization/antisymmetrization procedure}
\label{appB}
\begin{figure}
\includegraphics[width=1\linewidth, height=7cm]{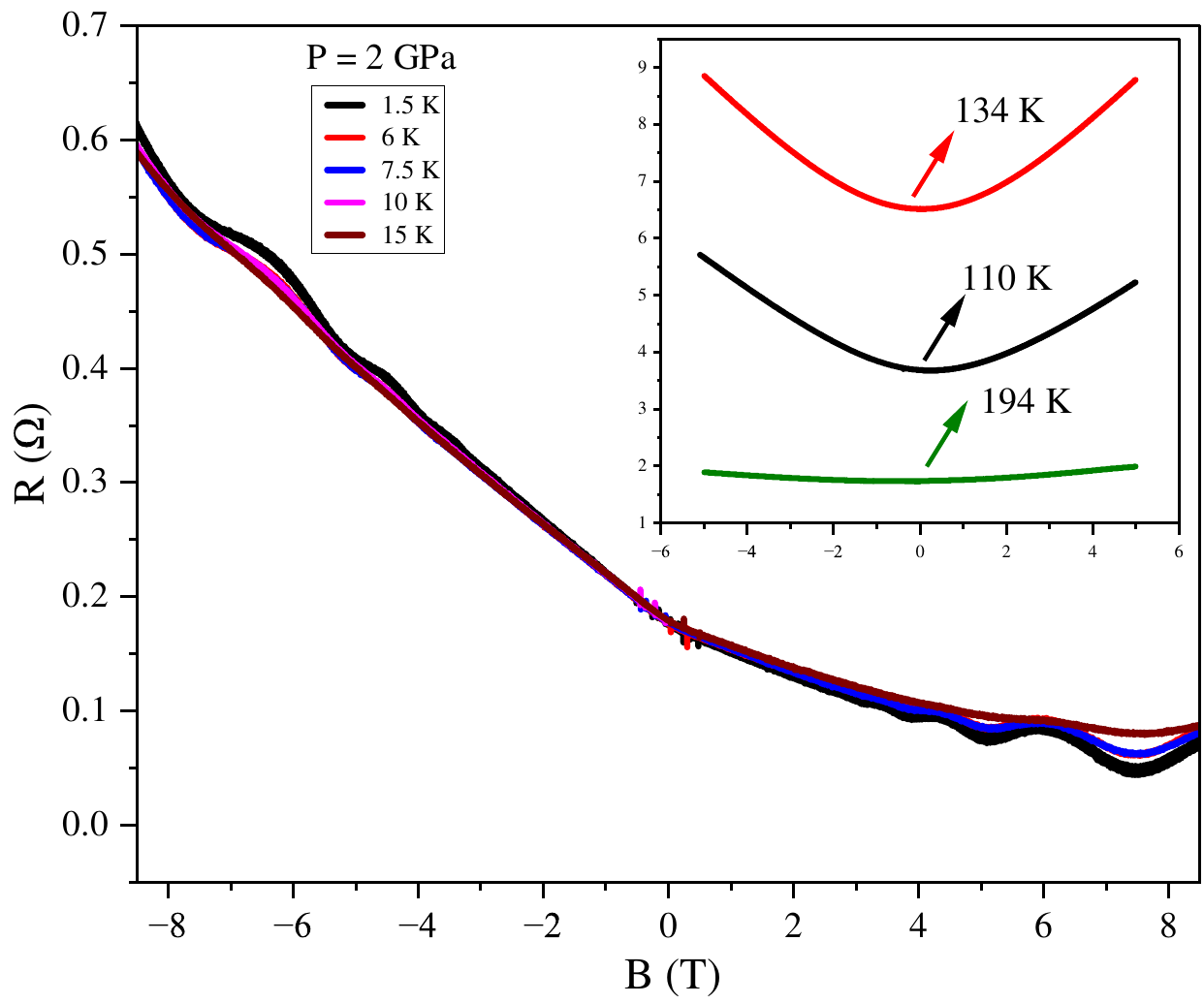}% Here is how to import EPS art
\caption{\label{fig:wide}(Appendix B) Raw magnetotransport data at a pressure of 2~GPa prior to symmetrization and antisymmetrization. The main panel shows data in the low-temperature regime (1.5--15~K), while the inset displays the high-temperature regime (134--194~K).}
\label{fig11}
\end{figure}

The raw data were measured under both positive and negative magnetic fields (Fig.~\ref{fig11}). From these measurements, the even and odd magnetic-field components were separated using the standard relations shown below~\ref{eq3}:
\begin{equation}
R_{xx}(B) = \frac{R(+B) + R(-B)}{2},
\end{equation}
\begin{equation}
R_{xy}(B) = \frac{R(+B) - R(-B)}{2}.
\label{eq3}
\end{equation}

In order to minimize possible hysteresis and instrumental effects, both forward and backward magnetic-field sweeps were averaged prior to the symmetrization and antisymmetrization procedures. Interpolation was applied to ensure that the resistance values at $+B$ and $-B$ correspond to identical magnetic field magnitudes, as required for accurate symmetrization and antisymmetrization.

\section{}
\subsection*{ Hall conductivity $\sigma_{xy}$ at different pressures along with fitting curves using the two-band model.}
\label{appC}
\begin{figure}
\includegraphics[width=1\linewidth, height=6cm]{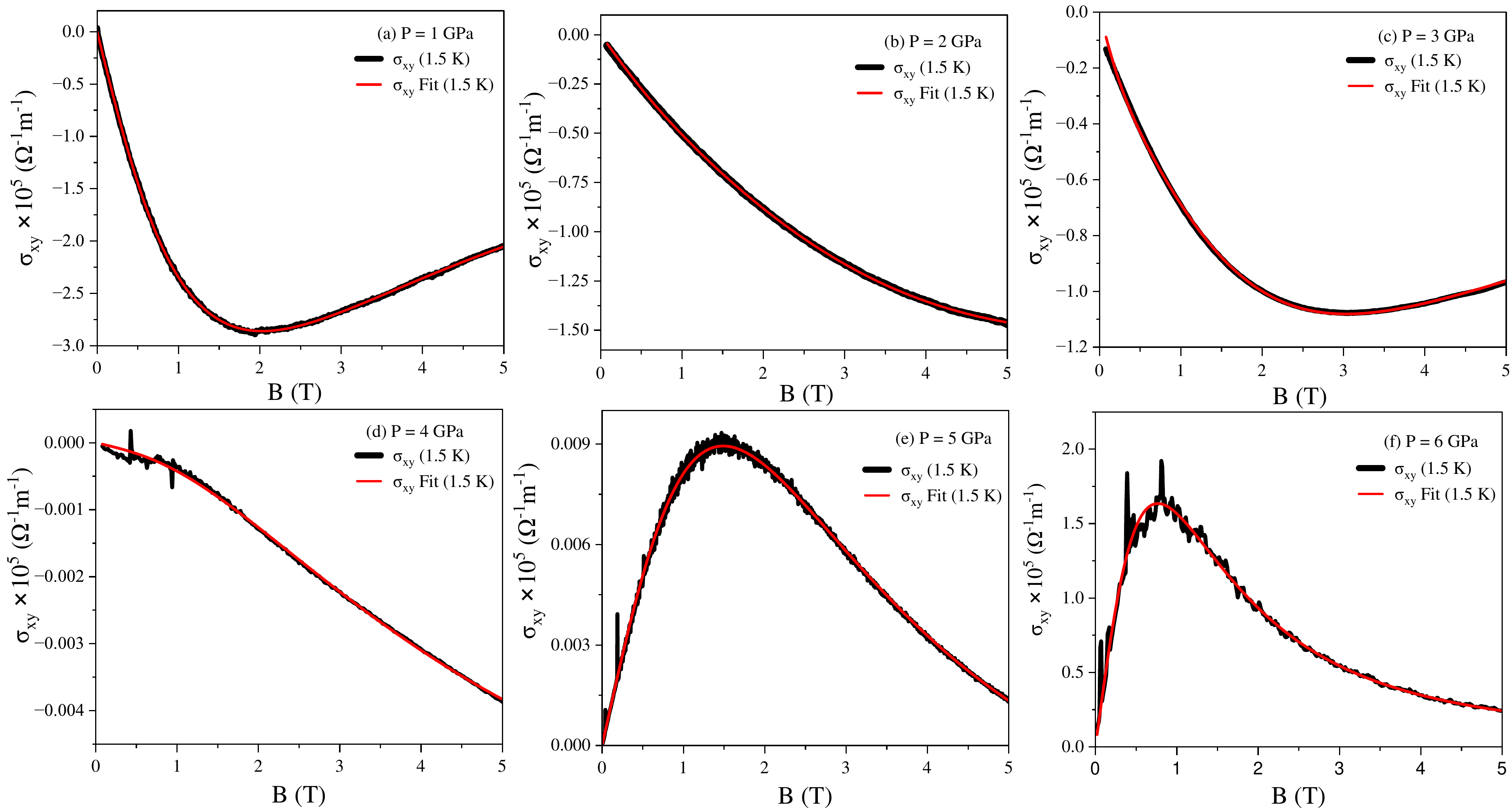}% Here is how to import EPS art
\caption{\label{fig:wide}(Appendix C) Hall conductivity $\sigma_{xy}$ at 1--6 ~GPa pressure along with the fitting curve shown in Eq. (2) }
\label{fig12}
\end{figure}
In Fig.~\ref{fig12}, we present the Hall conductivity $\sigma_{xy}$ at different pressures in the range of 1--6~GPa. The black curves represent the experimental data, while the red curves correspond to the fitting results obtained using Eq.~\ref{eq2}.

\section{}
\subsection*{ Band structure and Density of states at 0,2,3 and 5 GPa pressure}
\label{appD}
\begin{figure}
\includegraphics[width=1\linewidth, height=6cm]{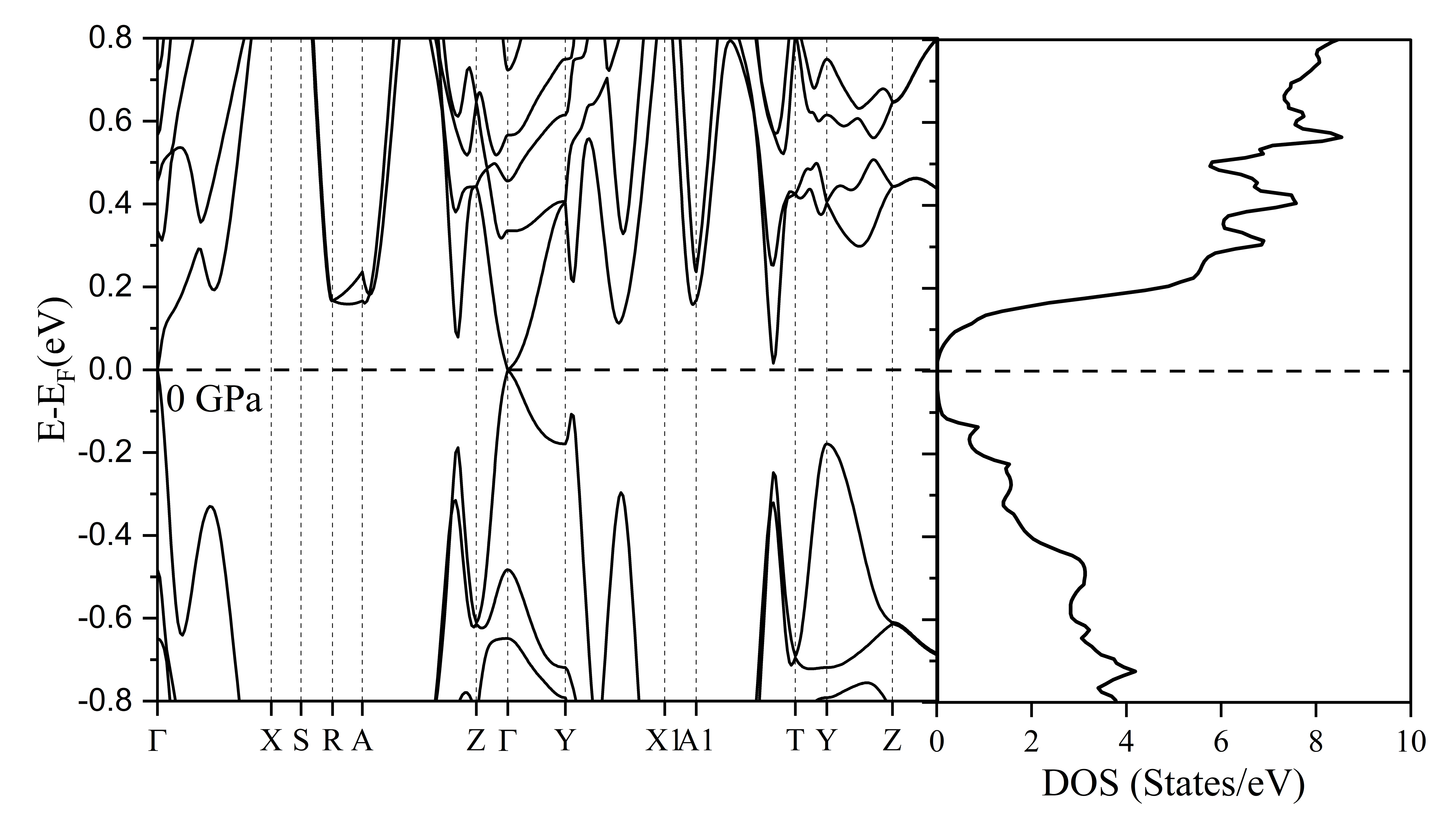}% Here is how to import EPS art
\caption{\label{fig:wide}(Appendix~\ref{appD}) Bands and DOS of ZrTe$_5$ at ambient pressure. Reproduced from our recent work \cite{Mishra2025}.}
\label{fig13}
\end{figure}

\begin{figure}
\includegraphics[width=1\linewidth, height=6cm]{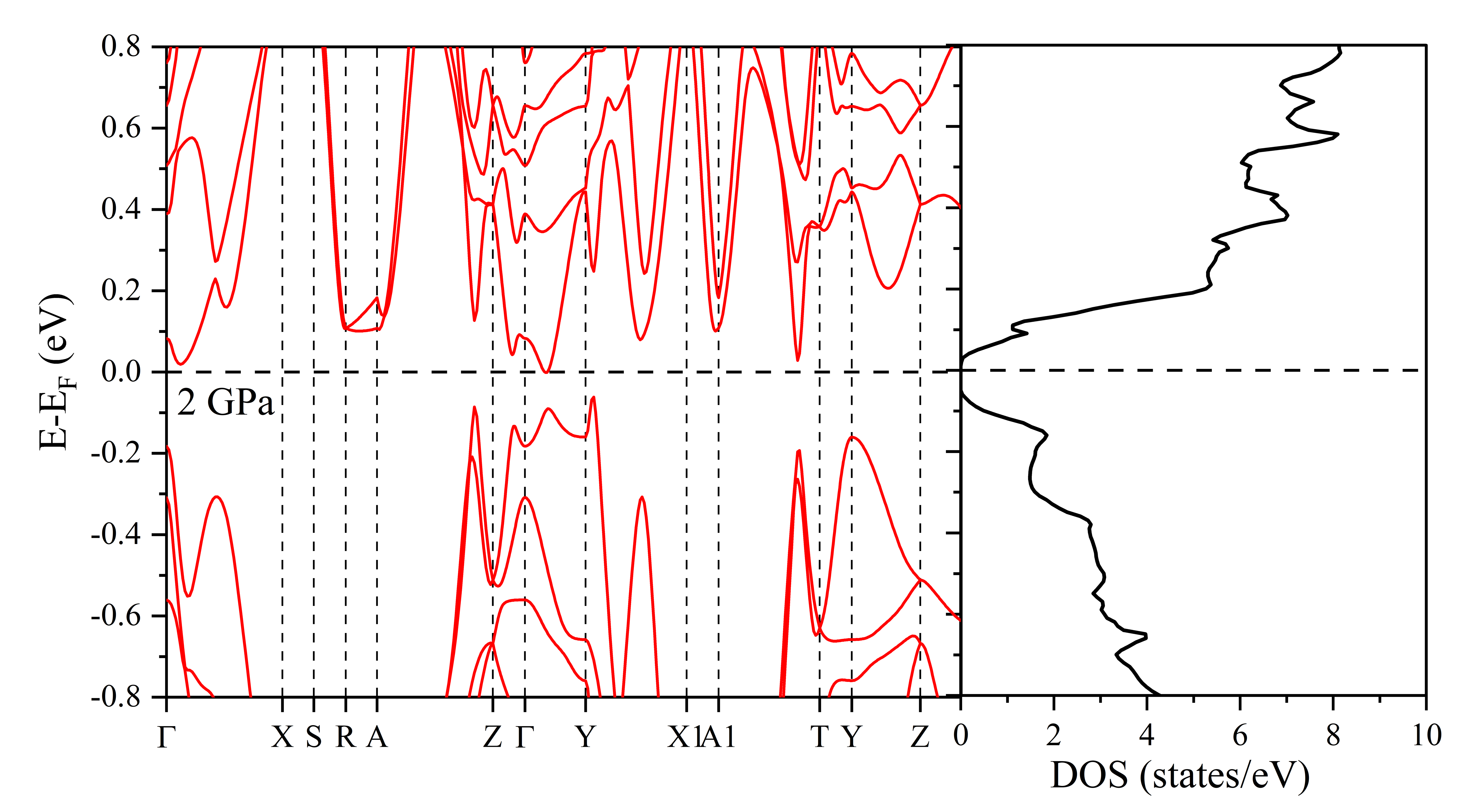}% Here is how to import EPS art
\caption{\label{fig:wide}(Appendix~\ref{appD}) Bands and DOS of ZrTe$_5$ at 2 GPa pressure. Reproduced from our recent work \cite{Mishra2025}.}
\label{fig14}
\end{figure}

\begin{figure}
\includegraphics[width=1\linewidth, height=6cm]{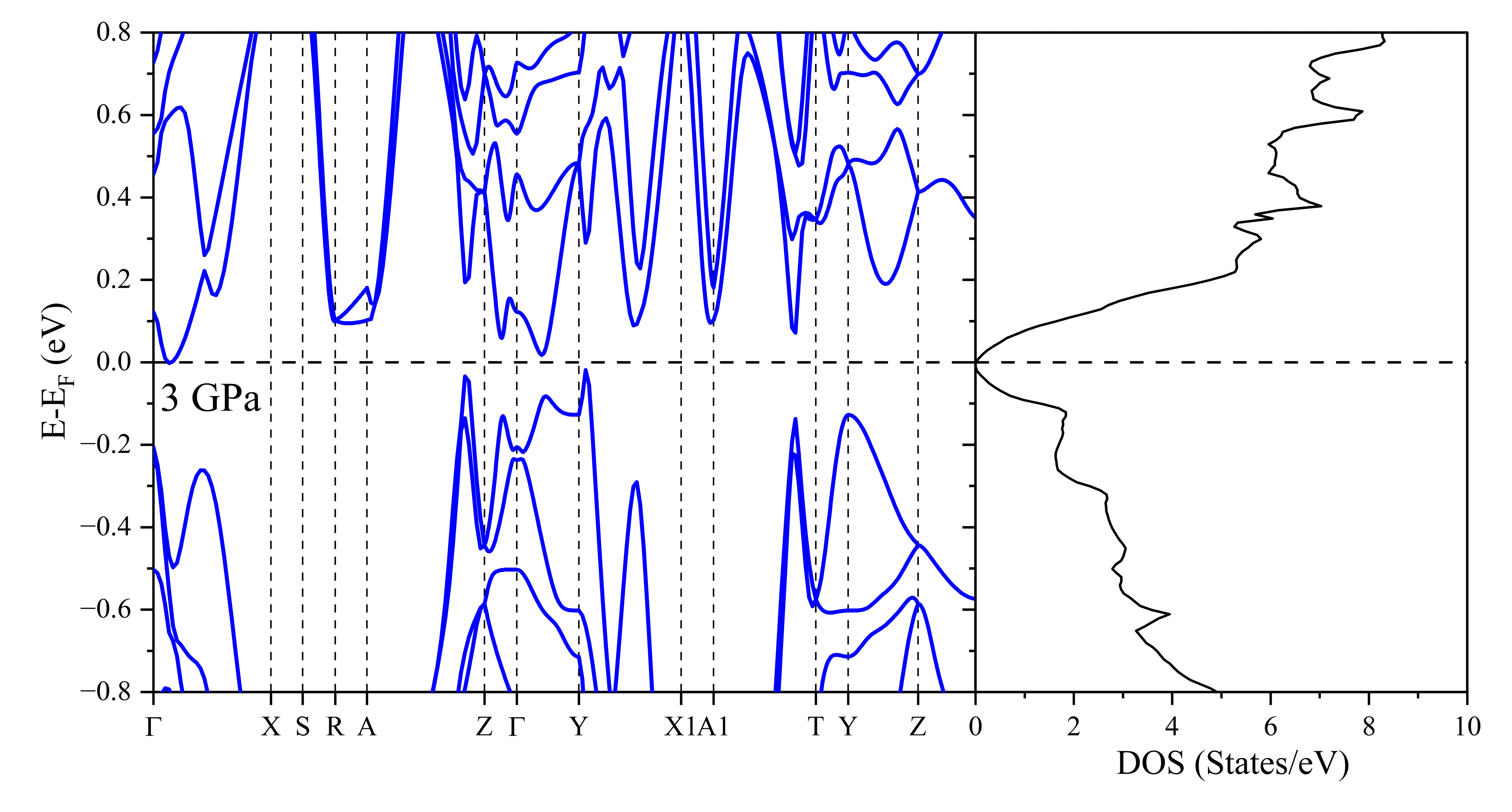}% Here is how to import EPS art
\caption{\label{fig:wide}(Appendix~\ref{appD}) Bands and DOS of ZrTe$_5$ at 3 GPa pressure.}
\label{fig15}
\end{figure}

In order to provide a comprehensive understanding of the effect of pressure on electronic properties of ZrTe$_5$, we present the band structure and density of states (DOS) at ambient, 2,3 and 5 GPa pressures, respectively. The Figure~\ref{fig13} $\&$~\ref{fig14} has been reproduced from our recent study \cite{Mishra2025}. It is clear form Figure~\ref{fig13}, that at ambient pressure there is Dirac-like band dispersion at $\Gamma$ point. As we increase pressure, the Fermi surface undergoes significant modification [Figure (14-16)]. A pronounced band-wrapping can be observed near the $\Gamma$, Z and Y points at 2 \& 3 GPa pressure [Figure~\ref{fig14}, Figure~\ref{fig15}]. At 5 GPa pressure two hole pockets and one electron pocket appear to cross the Fermi level.
\begin{figure}
\includegraphics[width=1\linewidth, height=6cm]{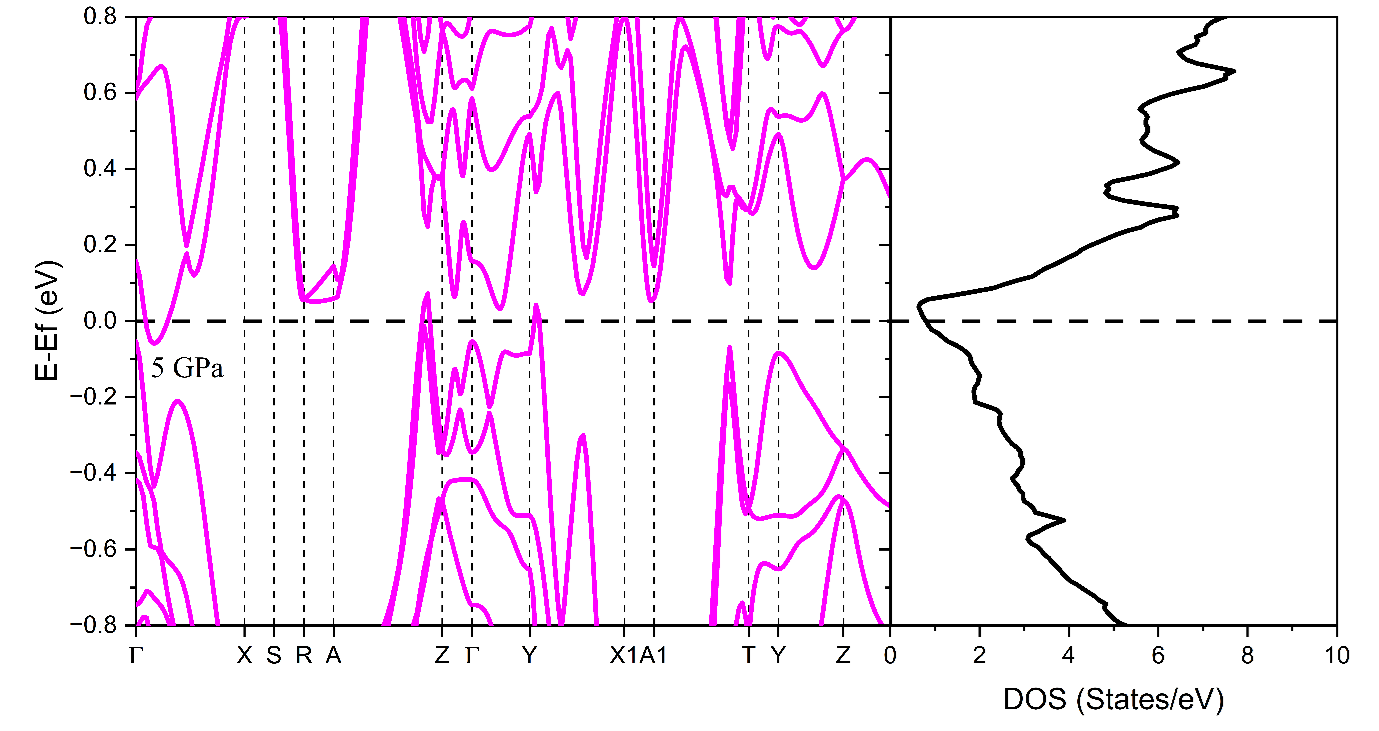}% Here is how to import EPS art
\caption{\label{fig:wide}(Appendix~\ref{appD}) Bands and DOS of ZrTe$_5$ at 5 GPa pressure.}
\label{fig16}
\end{figure} 
\FloatBarrier
\bibliography{references}% Produces the bibliography via BibTeX.

@article{Li2019 ,
  author   = {Li, Yupeng and Xu, Zhu-An},
  title    = {Exploring Topological Superconductivity in Topological Materials},
  journal  = {Advanced Quantum Technologies},
  volume   = {2},
  number   = {9},
  pages    = {1800112},
  year     = {2019},
  doi      = {10.1002/qute.201800112},
  url      = {https://advanced.onlinelibrary.wiley.com/doi/abs/10.1002/qute.201800112},
}

@article{Czhang2011,
  title = {Phase diagram of a pressure-induced superconducting state and its relation to the {Hall} coefficient of {Bi$_2$Te$_3$} single crystals},
  author = {Zhang, Chao and Sun, Liling and Chen, Zhaoyu and Zhou, Xingjiang and Wu, Qi and Yi, Wei and Guo, Jing and Dong, Xiaoli and Zhao, Zhongxian},
  journal = {Phys. Rev. B},
  volume = {83},
  issue = {14},
  pages = {140504},
  numpages = {4},
  year = {2011},
  month = {Apr},
  publisher = {American Physical Society},
  doi = {10.1103/PhysRevB.83.140504},
  url = {https://link.aps.org/doi/10.1103/PhysRevB.83.140504}
}

@article{Kirshenbaum2013,
  title = {Pressure-Induced Unconventional Superconducting Phase in the Topological Insulator {Bi$_2$Se$_3$}},
  author = {Kirshenbaum, Kevin and Syers, P. S. and Hope, A. P. and Butch, N. P. and Jeffries, J. R. and Weir, S. T. and Hamlin, J. J. and Maple, M. B. and Vohra, Y. K. and Paglione, J.},
  journal = {Phys. Rev. Lett.},
  volume = {111},
  issue = {8},
  pages = {087001},
  numpages = {5},
  year = {2013},
  month = {Aug},
  publisher = {American Physical Society},
  doi = {10.1103/PhysRevLett.111.087001},
  url = {https://link.aps.org/doi/10.1103/PhysRevLett.111.087001}
}

@article{Zhu2013,
  author    = {Zhu, J. and Zhang, J. L. and Kong, P. P. and Zhang, S. J. and Yu, X. H. and Zhu, J. L. and Liu, Q. Q. and Li, X. and Yu, R. C. and Ahuja, R. and Yang, W. G. and Shen, G. Y. and Mao, H. K. and Weng, H. M. and Dai, X. and Fang, Z. and Zhao, Y. S. and Jin, C. Q.},
  title     = {Superconductivity in Topological Insulator {Sb$_2$Te$_3$} Induced by Pressure},
  journal   = {Scientific Reports},
  volume    = {3},
  number    = {1},
  pages     = {2016},
  year      = {2013},
  doi       = {10.1038/srep02016},
  url       = {https://doi.org/10.1038/srep02016},
}

@article{JLZhang2011,
author = {J. L. Zhang  and S. J. Zhang  and H. M. Weng  and W. Zhang  and L. X. Yang  and Q. Q. Liu  and S. M. Feng  and X. C. Wang  and R. C. Yu  and L. Z. Cao  and L. Wang  and W. G. Yang  and H. Z. Liu  and W. Y. Zhao  and S. C. Zhang  and X. Dai  and Z. Fang  and C. Q. Jin },
title = {Pressure-induced superconductivity in topological parent compound {Bi$_2$Te$_3$}},
journal = {Proceedings of the National Academy of Sciences},
volume = {108},
number = {1},
pages = {24-28},
year = {2011},
doi = {10.1073/pnas.1014085108},
URL = {https://www.pnas.org/doi/abs/10.1073/pnas.1014085108},
}

@article{He2016,
  author    = {He, Lanpo and Jia, Yating and Zhang, Sijia and Hong, Xiaochen and Jin, Changqing and Li, Shiyan},
  title     = {Pressure-induced superconductivity in the three-dimensional topological {{Dirac}} semimetal {Cd$_3$As$_2$}},
  journal   = {npj Quantum Materials},
  volume    = {1},
  number    = {1},
  pages     = {16014},
  year      = {2016},
  doi       = {10.1038/npjquantmats.2016.14},
  url       = {https://doi.org/10.1038/npjquantmats.2016.14},
}

@article{Kang2015,
  author    = {Kang, Defen and Zhou, Yazhou and Yi, Wei and Yang, Chongli and Guo, Jing and Shi, Youguo and Zhang, Shan and Wang, Zhe and Zhang, Chao and Jiang, Sheng and Li, Aiguo and Yang, Ke and Wu, Qi and Zhang, Guangming and Sun, Liling and Zhao, Zhongxian},
  title     = {Superconductivity emerging from a suppressed large magnetoresistant state in tungsten ditelluride},
  journal   = {Nature Communications},
  volume    = {6},
  number    = {1},
  pages     = {7804},
  year      = {2015},
  doi       = {10.1038/ncomms8804},
  url       = {https://doi.org/10.1038/ncomms8804},
}

@article{Qi2016,
  author    = {Qi, Yanpeng and Naumov, Pavel G. and Ali, Mazhar N. and Rajamathi, Catherine R. and Schnelle, Walter and Barkalov, Oleg and Hanfland, Michael and Wu, Shu-Chun and Shekhar, Chandra and Sun, Yan and Süß, Vicky and Schmidt, Marcus and Schwarz, Ulrich and Pippel, Eckhard and Werner, Peter and Hillebrand, Reinald and Förster, Tobias and Kampert, Erik and Parkin, Stuart and Cava, R. J. and Felser, Claudia and Yan, Binghai and Medvedev, Sergey A.},
  title     = {Superconductivity in {{Weyl}} semimetal candidate {MoTe$_2$}},
  journal   = {Nature Communications},
  volume    = {7},
  number    = {1},
  pages     = {11038},
  year      = {2016},
  doi       = {10.1038/ncomms11038},
  url       = {https://doi.org/10.1038/ncomms11038},
}

@article{Kob2015,
  title = {Topological Superconductivity in {{Dirac} Semimetals}},
  author = {Kobayashi, Shingo and Sato, Masatoshi},
  journal = {Phys. Rev. Lett.},
  volume = {115},
  issue = {18},
  pages = {187001},
  numpages = {5},
  year = {2015},
  month = {Oct},
  publisher = {American Physical Society},
  doi = {10.1103/PhysRevLett.115.187001},
  url = {https://link.aps.org/doi/10.1103/PhysRevLett.115.187001}
}

@article{Li2018,
  author    = {Li, Yupeng and An, Chao and Hua, Chenqiang and Chen, Xuliang and Zhou, Yonghui and Zhou, Ying and Zhang, Ranran and Park, Changyong and Wang, Zhen and Lu, Yunhao and Zheng, Yi and Yang, Zhaorong and Xu, Zhu-An},
  title     = {Pressure-induced superconductivity in topological semimetal {NbAs$_2$}},
  journal   = {npj Quantum Materials},
  volume    = {3},
  number    = {1},
  pages     = {58},
  year      = {2018},
  doi       = {10.1038/s41535-018-0132-1},
  url       = {https://doi.org/10.1038/s41535-018-0132-1},
}

@article{Nayak2017,
  author    = {Nayak, Jayita and Wu, Shu-Chun and Kumar, Nitesh and Shekhar, Chandra and Singh, Sanjay and Fink, Jörg and Rienks, Emile E. D. and Fecher, Gerhard H. and Parkin, Stuart S. P. and Yan, Binghai and Felser, Claudia},
  title     = {Multiple {{Dirac}} cones at the surface of the topological metal {LaBi}},
  journal   = {Nature Communications},
  volume    = {8},
  number    = {1},
  pages     = {13942},
  year      = {2017},
  doi       = {10.1038/ncomms13942},
  url       = {https://doi.org/10.1038/ncomms13942},
}

@article{Lv2017,
  author    = {Lv, B. Q. and Feng, Z.-L. and Xu, Q.-N. and Gao, X. and Ma, J.-Z. and Kong, L.-Y. and Richard, P. and Huang, Y.-B. and Strocov, V. N. and Fang, C. and Weng, H.-M. and Shi, Y.-G. and Qian, T. and Ding, H.},
  title     = {Observation of three-component fermions in the topological semimetal molybdenum phosphide},
  journal   = {Nature},
  volume    = {546},
  number    = {7660},
  pages     = {627--631},
  year      = {2017},
  doi       = {10.1038/nature22390},
  url       = {https://doi.org/10.1038/nature22390},
  issn      = {1476-4687}
}

@article{Chi2018,
  author  = {Chi, Zhenhua and Chen, Xuliang and An, Chao and Yang, Liuxiang and Zhao, Jinggeng and Feng, Zili and Zhou, Yonghui and Zhou, Ying and Gu, Chuanchuan and Zhang, Bowen and Yuan, Yifang and Kenney-Benson, Curtis and Yang, Wenge and Wu, Gang and Wan, Xiangang and Shi, Youguo and Yang, Xiaoping and Yang, Zhaorong},
  title   = {Pressure-induced superconductivity in {MoP}},
  journal = {npj Quantum Materials},
  year    = {2018},
  volume  = {3},
  number  = {1},
  pages   = {28},
  doi     = {10.1038/s41535-018-0102-7},
  url     = {https://doi.org/10.1038/s41535-018-0102-7},
}

@article{Weng2014,
  title = {Transition-Metal Pentatelluride $\mathrm{ZrTe}{}_{5}$ and $\mathrm{HfTe}{}_{5}$: A Paradigm for Large-Gap Quantum Spin Hall Insulators},
  author = {Weng, Hongming and Dai, Xi and Fang, Zhong},
  journal = {Phys. Rev. X},
  volume = {4},
  issue = {1},
  pages = {011002},
  numpages = {8},
  year = {2014},
  month = {Jan},
  publisher = {American Physical Society},
  doi = {10.1103/PhysRevX.4.011002},
  url = {https://link.aps.org/doi/10.1103/PhysRevX.4.011002}
}

@article{Okada1980,
  author    = {Okada, Shigeto and Sambongi, Takashi and Ido, Masayuki},
  title     = {Giant resistivity anomaly in {ZrTe$_5$}},
  journal   = {Journal of the Physical Society of Japan},
  year      = {1980},
  volume    = {49},
  number    = {2},
  pages     = {839--840},
  doi       = {10.1143/JPSJ.49.839},
  issn      = {0031-9015},
  publisher = {The Physical Society of Japan},
}

@article{Jones1982,
title = {Thermoelectric power of {HfTe$_5$} and {ZrTe$_5$}},
journal = {Solid State Communications},
volume = {42},
number = {11},
pages = {793-798},
year = {1982},
issn = {0038-1098},
doi = {https://doi.org/10.1016/0038-1098(82)90008-4},
url = {https://www.sciencedirect.com/science/article/pii/0038109882900084},
author = {T.E. Jones and W.W. Fuller and T.J. Wieting and F. Levy},
}

@article{Li2016,
  author    = {Li, Qiang and Kharzeev, Dmitri E. and Zhang, Cheng and Huang, Yuan and Pletikosi{\'c}, I. and Fedorov, A. V. and Zhong, R. D. and Schneeloch, J. A. and Gu, G. D. and Valla, T.},
  title     = {Chiral magnetic effect in {ZrTe$_5$}},
  journal   = {Nature Physics},
  year      = {2016},
  volume    = {12},
  number    = {6},
  pages     = {550--554},
  doi       = {10.1038/nphys3648},
  url       = {https://doi.org/10.1038/nphys3648}
}

@article{Liang2018,
  author    = {Liang, Tian and Lin, Jingjing and Gibson, Quinn and Kushwaha, Satya and Liu, Minhao and Wang, Wudi and Xiong, Hongyu and Sobota, Jonathan A. and Hashimoto, Makoto and Kirchmann, Patrick S. and Shen, Zhi-Xun and Cava, R. J. and Ong, N. P.},
  title     = {Anomalous {Hall} effect in {ZrTe$_5$}},
  journal   = {Nature Physics},
  year      = {2018},
  volume    = {14},
  number    = {5},
  pages     = {451--455},
  doi       = {10.1038/s41567-018-0078-z},
  issn      = {1745-2481},
  url       = {https://doi.org/10.1038/s41567-018-0078-z}
}

@article{zhang2019,
  title = {Anomalous Thermoelectric Effects of $\mathrm{ZrTe_5}$ in and beyond the Quantum Limit},
  author = {Zhang, J. L. and Wang, C. M. and Guo, C. Y. and Zhu, X. D. and Zhang, Y. and Yang, J. Y. and Wang, Y. Q. and Qu, Z. and Pi, L. and Lu, Hai-Zhou and Tian, M. L.},
  journal = {Phys. Rev. Lett.},
  volume = {123},
  issue = {19},
  pages = {196602},
  numpages = {6},
  year = {2019},
  month = {Nov},
  publisher = {American Physical Society},
  doi = {10.1103/PhysRevLett.123.196602},
  url = {https://link.aps.org/doi/10.1103/PhysRevLett.123.196602}
}

@article{Zhang2020,
  author    = {Zhang, Wenjie and Wang, Peipei and Skinner, Brian and Bi, Ran and Kozii, Vladyslav and Cho, Chang-Woo and Zhong, Ruidan and Schneeloch, John and Yu, Dapeng and Gu, Genda and Fu, Liang and Wu, Xiaosong and Zhang, Liyuan},
  title     = {Observation of a thermoelectric {Hall} plateau in the extreme quantum limit},
  journal   = {Nature Communications},
  year      = {2020},
  volume    = {11},
  number    = {1},
  pages     = {1046},
  doi       = {10.1038/s41467-020-14819-7},
}

@article{Wang2022,
  title = {Gigantic Magnetochiral Anisotropy in the Topological Semimetal $\mathrm{ZrTe_5}$},
  author = {Wang, Yongjian and Legg, Henry F. and B\"omerich, Thomas and Park, Jinhong and Biesenkamp, Sebastian and Taskin, A. A. and Braden, Markus and Rosch, Achim and Ando, Yoichi},
  journal = {Phys. Rev. Lett.},
  volume = {128},
  issue = {17},
  pages = {176602},
  numpages = {5},
  year = {2022},
  month = {Apr},
  publisher = {American Physical Society},
  doi = {10.1103/PhysRevLett.128.176602},
  url = {https://link.aps.org/doi/10.1103/PhysRevLett.128.176602}
}

@article{Fan2017,
  author    = {Fan, Zongjian and Liang, Qi-Feng and Chen, Y. B. and Yao, Shu-Hua and Zhou, Jian},
  title     = {Transition between strong and weak topological insulator in {ZrTe$_5$} and {HfTe$_5$}},
  journal   = {Scientific Reports},
  year      = {2017},
  volume    = {7},
  number    = {1},
  pages     = {45667},
  doi       = {10.1038/srep45667},
  issn      = {2045-2322},
  url       = {https://doi.org/10.1038/srep45667}
}

@article{Liu2017,
  author    = {Liu, Y. and Long, Y. J. and Zhao, L. X. and Nie, S. M. and Zhang, S. J. and Weng, Y. X. and Jin, M. L. and Li, W. M. and Liu, Q. Q. and Long, Y. W. and Yu, R. C. and Gu, C. Z. and Sun, F. and Yang, W. G. and Mao, H. K. and Feng, X. L. and Li, Q. and Zheng, W. T. and Weng, H. M. and Dai, X. and Fang, Z. and Chen, G. F. and Jin, C. Q.},
  title     = {Superconductivity in {HfTe$_5$} across weak to strong topological insulator transition induced via pressures},
  journal   = {Scientific Reports},
  year      = {2017},
  volume    = {7},
  number    = {1},
  pages     = {44367},
  doi       = {10.1038/srep44367},
  issn      = {2045-2322},
  url       = {https://doi.org/10.1038/srep44367}
}

@article{KK2024,
  author    = {Zolt{\'a}n Kov{\'a}cs-Krausz and D{\'a}niel Nagy and Albin M{\'a}rffy and Bogdan Karpiak and Zolt{\'a}n Tajkov and L{\'a}szl{\'o} Oroszl{\'a}ny and J{\'a}nos Koltai and P{\'e}ter Nemes-Incze and Saroj P. Dash and P{\'e}ter Makk and Szabolcs Csonka and Endre T{\'o}v{\'a}ri},
  title     = {Signature of pressure-induced topological phase transition in {ZrTe\textsubscript{5}}},
  journal   = {npj Quantum Materials},
  volume    = {9},
  number    = {1},
  pages     = {76},
  year      = {2024},
  month     = {oct},
  doi       = {10.1038/s41535-024-00679-7},
  url       = {https://doi.org/10.1038/s41535-024-00679-7},
}

@article{Zhou2016,
  author    = {Y. Zhou and J. Wu and W. Ning and N. Li and Y. Du and X. Chen and R. Zhang and Z. Chi and X. Wang and X. Zhu and P. Lu and C. Ji and X. Wan and Z. Yang and J. Sun and W. Yang and M. Tian and Y. Zhang and H. Mao},
  title     = {Pressure-induced superconductivity in a three-dimensional topological material {ZrTe\textsubscript{5}}},
  journal   = {Proceedings of the National Academy of Sciences of the United States of America},
  volume    = {113},
  number    = {11},
  pages     = {2904--2909},
  year      = {2016},
  doi       = {10.1073/pnas.1601262113},
  url       = {https://doi.org/10.1073/pnas.1601262113}
}

@article{Mishra2025,
  title = {Effect of pressure on the transport properties and thermoelectric performance of {Dirac} semimetal $\mathrm{ZrTe_5}$},
  author = {Mishra, Sanskar and Singh, Nagendra and Gangwar, V. K. and Walia, Rajan and Kumar, Manindra and Singh, Udai Bhan and Saini, Deepash Sekhar and Sun, Jianping and Chen, Genfu and Bhoi, Dilip and Chatterjee, Sandip and Uwatoko, Yoshiya and Cheng, Jinguang and Shahi, Prashant},
  journal = {Phys. Rev. B},
  volume = {112},
  issue = {16},
  pages = {165140},
  numpages = {14},
  year = {2025},
  month = {Oct},
  publisher = {American Physical Society},
  doi = {10.1103/53lr-gc9w},
  url = {https://link.aps.org/doi/10.1103/53lr-gc9w}
}

@article{dis2017,
  title = {Disruption of the Accidental {Dirac} Semimetal State in $\mathrm{ZrTe_5}$ under Hydrostatic Pressure},
  author = {Zhang, J. L. and Guo, C. Y. and Zhu, X. D. and Ma, L. and Zheng, G. L. and Wang, Y. Q. and Pi, L. and Chen, Y. and Yuan, H. Q. and Tian, M. L.},
  journal = {Phys. Rev. Lett.},
  volume = {118},
  issue = {20},
  pages = {206601},
  numpages = {5},
  year = {2017},
  month = {May},
  publisher = {American Physical Society},
  doi = {10.1103/PhysRevLett.118.206601},
  url = {https://link.aps.org/doi/10.1103/PhysRevLett.118.206601}
}

@article{Qi_2016,
  title = {Pressure-driven superconductivity in the transition-metal pentatelluride $\mathrm{HfT}{\mathrm{e}}_{5}$},
  author = {Qi, Yanpeng and Shi, Wujun and Naumov, Pavel G. and Kumar, Nitesh and Schnelle, Walter and Barkalov, Oleg and Shekhar, Chandra and Borrmann, Horst and Felser, Claudia and Yan, Binghai and Medvedev, Sergey A.},
  journal = {Phys. Rev. B},
  volume = {94},
  issue = {5},
  pages = {054517},
  numpages = {9},
  year = {2016},
  month = {Aug},
  publisher = {American Physical Society},
  doi = {10.1103/PhysRevB.94.054517},
  url = {https://link.aps.org/doi/10.1103/PhysRevB.94.054517}
}

@article{Ana2020,
  title = {Probing intraband excitations in {${\mathrm{ZrTe}}_{5}$}: A high-pressure infrared and transport study},
  author = {Santos-Cottin, D. and Padlewski, M. and Martino, E. and David, S. Ben and Le Mardel\'e, F. and Capitani, F. and Borondics, F. and Bachmann, M. D. and Putzke, C. and Moll, P. J. W. and Zhong, R. D. and Gu, G. D. and Berger, H. and Orlita, M. and Homes, C. C. and Rukelj, Z. and Akrap, Ana},
  journal = {Phys. Rev. B},
  volume = {101},
  issue = {12},
  pages = {125205},
  numpages = {10},
  year = {2020},
  month = {Mar},
  publisher = {American Physical Society},
  doi = {10.1103/PhysRevB.101.125205},
  url = {https://link.aps.org/doi/10.1103/PhysRevB.101.125205}
}

@article{Shahi2018,
  title = {Bipolar Conduction as the Possible Origin of the Electronic Transition in Pentatellurides: Metallic vs Semiconducting Behavior},
  author = {Shahi, P. and Singh, D. J. and Sun, J. P. and Zhao, L. X. and Chen, G. F. and Lv, Y. Y. and Li, J. and Yan, J.-Q. and Mandrus, D. G. and Cheng, J.-G.},
  journal = {Phys. Rev. X},
  volume = {8},
  issue = {2},
  pages = {021055},
  numpages = {13},
  year = {2018},
  month = {May},
  publisher = {American Physical Society},
  doi = {10.1103/PhysRevX.8.021055},
  url = {https://link.aps.org/doi/10.1103/PhysRevX.8.021055}
}

@article{Cheng2018,
doi = {10.1088/1674-1056/27/7/077403},
url = {https://dx.doi.org/10.1088/1674-1056/27/7/077403},
year = {2018},
month = {jul},
publisher = {Chinese Physical Society and IOP Publishing Ltd},
volume = {27},
number = {7},
pages = {077403},
author = {Cheng, Jin-Guang and Wang, Bo-Sen and Sun, Jian-Ping and Uwatoko, Yoshiya},
title = {Cubic anvil cell apparatus for high-pressure and low-temperature physical property measurements*},
journal = {Chinese Physics B},
}

@article{Uwatoko2008,
  title     = {Development of an ultra-compact cubic-anvil pressure apparatus for ultralow temperatures},
  author    = {Yoshiya Uwatoko and Kazuyuki Matsubayashi and Takehiko Matsumoto and Naofumi Aso and Masakazu Nishi and Tetsuya Fujiwara and Masato Hondo and Satoshi Tabata and Katsuhiro Takagi and Masashi Tawata and Hiroyuki Kagi},
  journal   = {High Pressure Research},
  volume    = {18},
  number    = {3},
  pages     = {230--236},
  year      = {2008},
  doi       = {10.4131/jshpreview.18.230},
  language  = {English},
}

@article{Gian2017,
doi = {10.1088/1361-648X/aa8f79},
url = {https://dx.doi.org/10.1088/1361-648X/aa8f79},
year = {2017},
month = {oct},
publisher = {IOP Publishing},
volume = {29},
number = {46},
pages = {465901},
author = {Giannozzi, P and Andreussi, O and Brumme, T and Bunau, O and Buongiorno Nardelli, M and Calandra, M and Car, R and Cavazzoni, C and Ceresoli, D and Cococcioni, M and Colonna, N and Carnimeo, I and Dal Corso, A and de Gironcoli, S and Delugas, P and DiStasio, R A and Ferretti, A and Floris, A and Fratesi, G and Fugallo, G and Gebauer, R and Gerstmann, U and Giustino, F and Gorni, T and Jia, J and Kawamura, M and Ko, H-Y and Kokalj, A and Küçükbenli, E and Lazzeri, M and Marsili, M and Marzari, N and Mauri, F and Nguyen, N L and Nguyen, H-V and Otero-de-la-Roza, A and Paulatto, L and Poncé, S and Rocca, D and Sabatini, R and Santra, B and Schlipf, M and Seitsonen, A P and Smogunov, A and Timrov, I and Thonhauser, T and Umari, P and Vast, N and Wu, X and Baroni, S},
title = {Advanced capabilities for materials modelling with Quantum ESPRESSO},
journal = {Journal of Physics: Condensed Matter},
}

@article{Per2017,
  title = {Generalized Gradient Approximation Made Simple},
  author = {Perdew, John P. and Burke, Kieron and Ernzerhof, Matthias},
  journal = {Phys. Rev. Lett.},
  volume = {77},
  issue = {18},
  pages = {3865--3868},
  numpages = {0},
  year = {1996},
  month = {Oct},
  publisher = {American Physical Society},
  doi = {10.1103/PhysRevLett.77.3865},
  url = {https://link.aps.org/doi/10.1103/PhysRevLett.77.3865}
}

@article{Grimme2010,
    author = {Grimme, Stefan and Antony, Jens and Ehrlich, Stephan and Krieg, Helge},
    title = {A consistent and accurate ab initio parametrization of density functional dispersion correction (DFT-D) for the 94 elements H-Pu},
    journal = {The Journal of Chemical Physics},
    volume = {132},
    number = {15},
    pages = {154104},
    year = {2010},
    month = {04},
    issn = {0021-9606},
    doi = {10.1063/1.3382344},
    url = {https://doi.org/10.1063/1.3382344},
}

@article{Zhang2017,
  title     = {Electronic evidence of temperature-induced Lifshitz transition and topological nature in {ZrTe\textsubscript{5}}},
  author    = {Zhang, Yan and Wang, Chenlu and Yu, Li and Liu, Guodong and Liang, Aiji and Huang, Jianwei and Nie, Simin and Sun, Xuan and Zhang, Yuxiao and Shen, Bing and Liu, Jing and Weng, Hongming and Zhao, Lingxiao and Chen, Genfu and Jia, Xiaowen and Hu, Cheng and Ding, Ying and Zhao, Wenjuan and Gao, Qiang and Li, Cong and He, Shaolong and Zhao, Lin and Zhang, Fengfeng and Zhang, Shenjin and Yang, Feng and Wang, Zhimin and Peng, Qinjun and Dai, Xi and Fang, Zhong and Xu, Zuyan and Chen, Chuangtian and Zhou, X. J.},
  journal   = {Nature Communications},
  volume    = {8},
  number    = {1},
  pages     = {15512},
  year      = {2017},
  doi       = {10.1038/ncomms15512},
  url       = {https://doi.org/10.1038/ncomms15512}
}

@article{Piva2024,
  title = {Importance of the semimetallic state for the quantum Hall effect in $\mathrm{HfTe_5}$},
  author = {Piva, M. M. and Wawrzy\ifmmode \acute{n}\else \'{n}\fi{}czak, R. and Kumar, Nitesh and Kutelak, L. O. and Lombardi, G. A. and dos Reis, R. D. and Felser, C. and Nicklas, M.},
  journal = {Phys. Rev. Mater.},
  volume = {8},
  issue = {4},
  pages = {L041202},
  numpages = {6},
  year = {2024},
  month = {Apr},
  publisher = {American Physical Society},
  doi = {10.1103/PhysRevMaterials.8.L041202},
  url = {https://link.aps.org/doi/10.1103/PhysRevMaterials.8.L041202}
}

@article{Gao2021,
title = {Effects of pressure on structural, electronic, optical, and mechanical properties of ZrTe5: A density functional theory study},
journal = {Physica B: Condensed Matter},
volume = {620},
pages = {413286},
year = {2021},
issn = {0921-4526},
doi = {https://doi.org/10.1016/j.physb.2021.413286},
url = {https://www.sciencedirect.com/science/article/pii/S0921452621004592},
author = {Juan Gao and Mi Zhong and Qi-Jun Liu and Bin Tang and Fu-Sheng Liu and Xiao-Juan Ma},
}

@article{Y2017,
title = {Temperature-induced Lifshitz transition in topological insulator candidate $\mathrm{HfTe_5}$},
journal = {Science Bulletin},
volume = {62},
number = {13},
pages = {950-956},
year = {2017},
issn = {2095-9273},
doi = {https://doi.org/10.1016/j.scib.2017.05.030},
url = {https://www.sciencedirect.com/science/article/pii/S2095927317302918},
author = {Yan Zhang and Chenlu Wang and Guodong Liu and Aiji Liang and Lingxiao Zhao and Jianwei Huang and Qiang Gao and Bing Shen and Jing Liu and Cheng Hu and Wenjuan Zhao and Genfu Chen and Xiaowen Jia and Li Yu and Lin Zhao and Shaolong He and Fengfeng Zhang and Shenjin Zhang and Feng Yang and Zhimin Wang and Qinjun Peng and Zuyan Xu and Chuangtian Chen and Xingjiang Zhou},
}

@article{Man2016,
  title = {Evidence for a Strong Topological Insulator Phase in $\mathrm{ZrTe_5}$},
  author = {Manzoni, G. and Gragnaniello, L. and Aut\`es, G. and Kuhn, T. and Sterzi, A. and Cilento, F. and Zacchigna, M. and Enenkel, V. and Vobornik, I. and Barba, L. and Bisti, F. and Bugnon, Ph. and Magrez, A. and Strocov, V. N. and Berger, H. and Yazyev, O. V. and Fonin, M. and Parmigiani, F. and Crepaldi, A.},
  journal = {Phys. Rev. Lett.},
  volume = {117},
  issue = {23},
  pages = {237601},
  numpages = {5},
  year = {2016},
  month = {Nov},
  publisher = {American Physical Society},
  doi = {10.1103/PhysRevLett.117.237601},
  url = {https://link.aps.org/doi/10.1103/PhysRevLett.117.237601}
}

@article{Osakabe_2008,
doi = {10.1143/JJAP.47.6544},
url = {https://doi.org/10.1143/JJAP.47.6544},
year = {2008},
month = {aug},
publisher = {},
volume = {47},
number = {8R},
pages = {6544},
author = {Osakabe, Toyotaka and Kakurai, Kazuhisa},
title = {Feasibility Tests on Pressure-Transmitting Media for Single-Crystal Magnetic Neutron Diffraction under High Pressure},
journal = {Japanese Journal of Applied Physics}
}

@article{Eguchi,
  title = {Robust scheme for magnetotransport analysis in topological insulators},
  author = {Eguchi, G. and Paschen, S.},
  journal = {Phys. Rev. B},
  volume = {99},
  issue = {16},
  pages = {165128},
  numpages = {9},
  year = {2019},
  month = {Apr},
  publisher = {American Physical Society},
  doi = {10.1103/PhysRevB.99.165128},
  url = {https://link.aps.org/doi/10.1103/PhysRevB.99.165128}
}
\end{document}